\documentclass[aps,prb,twocolumn,superscriptaddress,showpacs]{revtex4-1}

\usepackage{graphicx}
\usepackage{amsmath}
\usepackage{bm}
\usepackage{amsfonts}
\usepackage{amssymb}
\usepackage{xcolor}
\usepackage{natbib}

\begin{document}

\title{Spin dynamics and field-induced magnetic phase transition in the \\
honeycomb Kitaev magnet $\bm{\alpha}$-Li$_{2}$IrO$_{3}$}

\author{Sungkyun\ Choi}
\altaffiliation[Present address: ] {Rutgers Center for Emergent
Materials and Department of Physics and Astronomy, Rutgers
University, Piscataway, New Jersey 08854, USA}
\affiliation{Clarendon Laboratory, University of Oxford Physics
Department, Parks Road, Oxford, OX1 3PU, United Kingdom}
\author{S.\ Manni}
\altaffiliation[Present address: ] {
Indian Institute of Technology Palakkad, Kerala, India}
\affiliation{EP VI, Center for Electronic Correlations and
Magnetism, Augsburg University, D-86159 Augsburg, Germany}
\author{J.\ Singleton}
\affiliation{National High Magnetic Field Laboratory MPA-NHMFL,
TA-35, MS-E536 Los Alamos National Laboratory, Los Alamos, NM
87545, USA}
\author{C.V. Topping}
\affiliation{National High Magnetic Field Laboratory MPA-NHMFL,
TA-35, MS-E536 Los Alamos National Laboratory, Los Alamos, NM
87545, USA}
\affiliation{Clarendon Laboratory, University of Oxford Physics
Department, Parks Road, Oxford, OX1 3PU, United Kingdom}
\author{T.\ Lancaster}
\affiliation{Department of Physics, Durham University, South Road,
Durham, DH1 3LE, United Kingdom}
\author{S.~J.~Blundell}
\affiliation{Clarendon Laboratory, University of Oxford Physics
Department, Parks Road, Oxford, OX1 3PU, United Kingdom}
\author{D.\ T.\  Adroja}
\affiliation{ISIS Facility, Rutherford Appleton Laboratory-STFC,
Chilton, Didcot, OX11 0QX, United Kingdom}
\author{V. Zapf}
\affiliation{National High Magnetic Field Laboratory MPA-NHMFL,
TA-35, MS-E536 Los Alamos National Laboratory, Los Alamos, NM
87545, USA}
\author{P.\ Gegenwart}
\affiliation{EP VI, Center for Electronic Correlations and
Magnetism, Augsburg University, D-86159 Augsburg, Germany}
\author{R.\ Coldea}
\affiliation{Clarendon Laboratory, University of Oxford Physics
Department, Parks Road, Oxford, OX1 3PU, United Kingdom}

%\date{\today}

\begin{abstract}
The layered honeycomb iridate $\alpha$-Li$_2$IrO$_3$ displays an incommensurate magnetic structure with counterrotating moments on nearest-neighbor sites, proposed to be stabilized by strongly-frustrated anisotropic Kitaev interactions between spin-orbit entangled Ir$^{4+}$ magnetic moments. Here we report powder inelastic neutron scattering measurements that observe sharply dispersive low-energy magnetic excitations centered at the magnetic ordering wavevector, attributed to Goldstone excitations of the incommensurate order, as well as an additional intense mode above a gap $\Delta\simeq2.3$~meV. Zero-field muon-spin relaxation measurements show clear oscillations in the muon polarization below the N\'{e}el temperature $T_{\rm N}\simeq15$~K with a time-dependent profile consistent with bulk incommensurate long-range magnetism. Pulsed field magnetization measurements observe that only about half the saturation magnetization value is reached at the maximum field of 64~T. A clear anomaly near 25~T indicates a transition to a phase with reduced susceptibility. The transition field has a Zeeman energy comparable to the zero-field gapped mode, suggesting gap suppression as a possible mechanism for the field-induced transition.
\end{abstract}
\maketitle

\section{Introduction}
The cooperative magnetism of 4$d$ and 5$d$ transition metal ions with strong spin-orbit coupling is attracting much interest as a platform to potentially realize unconventional magnetic states stabilized by strong frustration effects from bond-dependent anisotropic couplings (Refs.~\onlinecite{Rau2016, Winter2016,Hermanns2018}). A canonical Hamiltonian in this context is the Kitaev model on the honeycomb lattice (Ref.~\onlinecite{kitaev}) with orthogonal moment components coupled via Ising interactions along the three bonds emerging out of each site. This leads to strong frustration effects that stabilize a quantum spin liquid ground state with exotic quasiparticles (Ref.~\onlinecite{knolle}). Potential hosts of Kitaev physics are tri-coordinated lattices of 5$d^5$ Ir$^{4+}$ or 4$d^5$ Ru$^{3+}$ ions inside edge-sharing octahedra, where spin-orbit entangled $J_{\rm eff}=1/2$ moments (stabilized by spin-orbit coupling and cubic crystal field) are expected to interact to leading order via Ising couplings (Ref.~\onlinecite{Jackeli:zz}). Candidate materials to realize such
interactions include the layered honeycomb Na$_2$IrO$_3$ (Ref.~\onlinecite{Gegenwart2010}), $\alpha$-RuCl$_3$ (Ref.~\onlinecite{Plumb2014}) and $\alpha$-Li$_2$IrO$_3$ (Ref.~\onlinecite{Yogesh}), as well as the three-dimensional structural polytypes $\beta$-(Ref.~\onlinecite{Takayama:beta:Li213}) and $\gamma$-Li$_2$IrO$_3$ (Ref.~\onlinecite{modic}) with hyperhoneycomb and stripyhoneycomb magnetic lattices, respectively. The current understanding is that all the above materials have strong Kitaev exchanges, but additional sub-leading interactions also present are sufficient to instead stabilize magnetic order: zigzag antiferromagnetism for Na$_2$IrO$_3$ (Refs.~\onlinecite{Liu2011,Na213_INS_2012,Ye2012,Hwan2015,Das2017}) and $\alpha$-RuCl$_3$ (Refs.~\onlinecite{Johnson2015,Banerjee2016}), and incommensurate counterrotating structures for the Li$_2$IrO$_3$ family (Refs.~\onlinecite{Steph:Li213,alun:beta:Li213,alun:gamma:Li213}). Promising avenues explored to suppress long-range magnetic order are hydrogen-intercalation in H$_3$LiIr$_2$O$_6$ (Ref.~\onlinecite{Takagi2017}) and high-pressure studies of $\beta$-Li$_2$IrO$_3$ (Refs.~\onlinecite{Veiga2017,Majumder2018,Takayama2017,bLIO_HP_Raman2018}) and $\gamma$-Li$_2$IrO$_3$ (Ref.~\onlinecite{Breznay2017}).

The most detailed spin dynamics studies are available for $\alpha$-RuCl$_3$, where inelastic neutron scattering experiments observe strong scattering continua indicative of large quantum fluctuations co-existing with magnetic order (Refs.~\onlinecite{Banerjee2017,Do2017}). Dispersive magnetic excitations have also been observed in Na$_2$IrO$_3$ via neutron scattering (Ref.~\onlinecite{Na213_INS_2012}) and resonant inelastic x-ray scattering (Ref.~\onlinecite{Na213_RIXS_GH_2013}) but the spin dynamics in the Li$_2$IrO$_3$ family has not been reported so far. Here we present first inelastic neutron scattering measurements of the spin dynamics in the $\alpha$ polytype. All three polytypes display closely related incommensurate counterrotating structures; counterrotation cannot be explained by Heisenberg-type interactions and provides direct evidence for the presence of dominant bond-dependent anisotropic couplings (Refs.~\onlinecite{Kimchi:abcLi213,Lee2015,Lee2016}). The excitations of incommensurate counterrotating structures are of fundamental conceptual interest as the counterrotation renders standard spin-wave approaches inapplicable and the theoretical spectrum is known only in a few special cases (Refs.~\onlinecite{Kimchi2016,Perkins2017}).

Another potential route to observe novel effects due to strong Kitaev interactions is in the behaviour in high applied magnetic fields. It is notable that $\alpha$-RuCl$_3$ (Refs.~\onlinecite{Johnson2015,Sears2017}) and $\beta$-Li$_2$IrO$_3$ (Refs.~\onlinecite{Takayama:beta:Li213,Ruiz2017}), and potentially also $\gamma$-Li$_2$IrO$_3$ (Ref.~\onlinecite{Modic2017}), show suppression of the spontaneous magnetic order at relatively low applied magnetic fields when compared to the dominant magnetic interaction strength, suggesting that strong quantum fluctuations, potentially enhanced by the proximity to a nearby spin-liquid phase in parameter space, play an important role in the suppression of the magnetic order. The mechanism of the field-induced transition and the properties above the critical field in those Kitev materials are currently attracting much interest, both experimentally as well as theoretically (Refs.~\onlinecite{BanerjeeB,WinterB,Lampen2018}).

Here we extend the investigation of the magnetic behavior of $\alpha$-Li$_2$IrO$_3$ by exploring the magnetic phase diagram up to 64~T. This reveals a field-induced transition near 25~T to another magnetic phase with reduced susceptibility and magnetization still much smaller than the expected saturated value. We also report measurements of the spin dynamics over a wide energy range using time-of-flight inelastic neutron scattering with an optimized setup to minimize neutron absorption. At low temperatures in the magnetically ordered phase we observe a clear dispersive inelastic magnetic signal centered at the magnetic
ordering wavevector, attributed to Goldstone mode fluctuations of the incommensurate magnetic order. In addition, we also find an intense gapped mode, which may be due to fluctuations out of the moment
rotation plane. The gapped mode energy is comparable to the Zeeman energy of the transition field observed in the pulsed-field magnetization data, suggesting gap suppression as a possible mechanism of the field-induced transition.

The paper is organized as follows. We first present in Secs.~\ref{sec:chi}-\ref{sec:muSR} magnetic susceptibility and muon spin relaxation measurements of powder $\alpha$-Li$_2$IrO$_3$ samples that confirm the presence of a sharp magnetic transition near 15~K to a well-defined, long-range magnetic order, which pervades the bulk of the samples. Sec.~\ref{sec:INS} shows measurements of the spin dynamics via powder inelastic neutron scattering, with the data parameterized in Sec.~\ref{sec:par} in terms of two magnetic excitations, a linearly-dispersive, gapless mode and an additional quadratic mode above a finite energy gap. Sec.~\ref{sec:mag} presents magnetization measurements in pulsed fields, which observe a clear anomaly indicative of a magnetic transition at a critical field of Zeeman energy comparable to the zero-field gapped mode energy. Sec.~\ref{sec:disc} discusses the results in the context of the expected mean-field phase diagram of incommensurate spiral-ordered magnets in applied field and the empirical magnetic phase diagram of other Kitaev magnets. Finally, Sec.~\ref{sec:sum} summarises the main results and conclusions.

\section{Magnetic susceptibility}
\label{sec:chi} 
\begin{figure}[tbh]
\begin{center}
% \hspace{1cm}
\includegraphics[width=\linewidth]{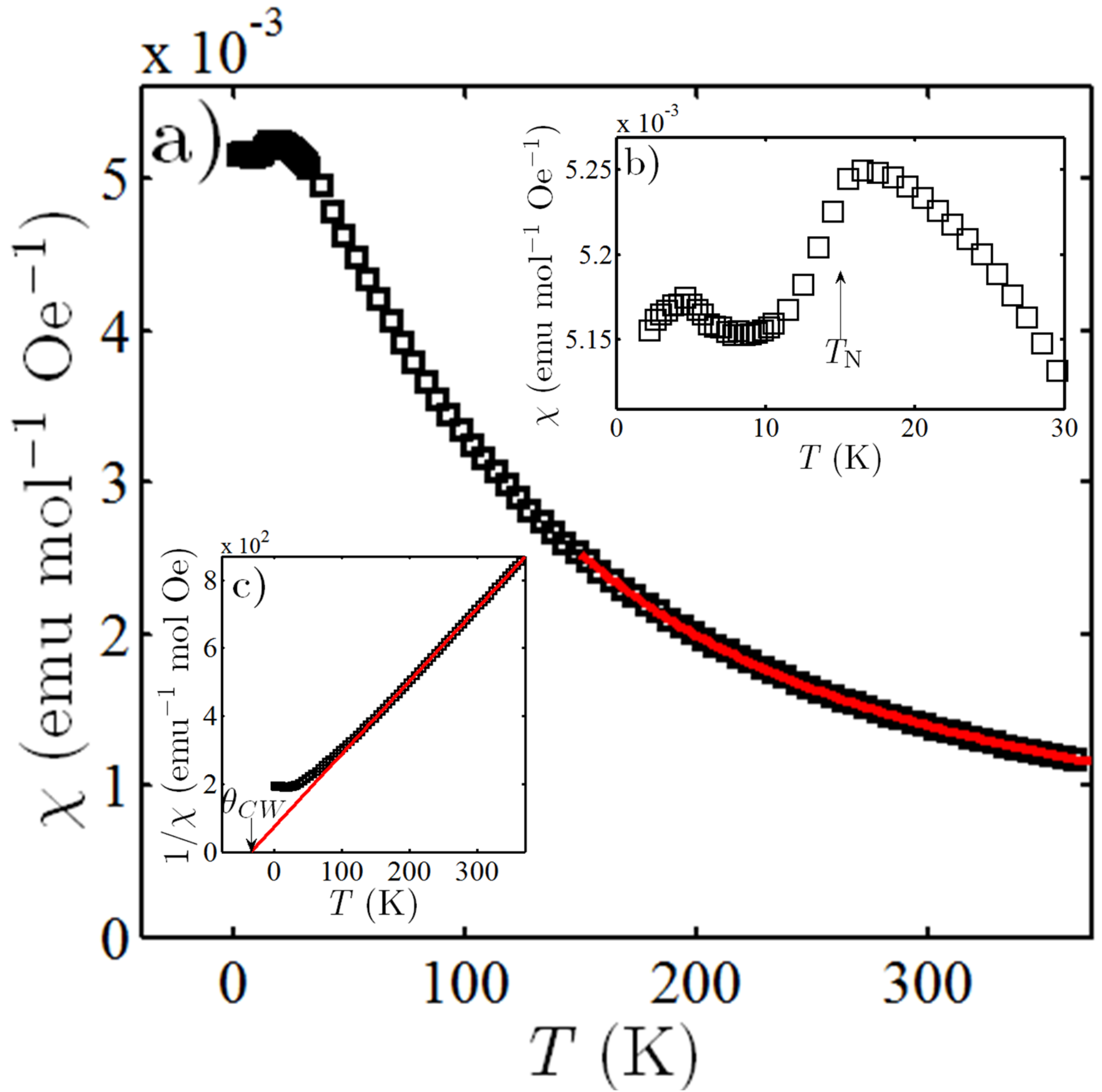}
\end{center} \caption {(color online) a) Temperature dependence of the magnetic susceptibility in powder samples of $\alpha$-Li$_2$IrO$_3$ (zero-field cooled, measurement field $\mu_0H=1000$~Oe). The red solid line is a fit to a Curie-Weiss form as discussed in the text. b) Expansion of the low-temperature region showing a clear anomaly at the magnetic ordering transition near 15~K. c) Inverse susceptibility fitted to a Curie-Weiss form (red solid line) over the range 150-370~K.} \label{exc_Li213:order:sus}
\end{figure}

Magnetic susceptibility measurements were performed using a Quantum Design SQUID magnetometer in the Clarendon Laboratory in Oxford on a fine powder sample ($56.5$~mg) of $\alpha$-Li$_{2}$IrO$_{3}$ synthesized as described elsewhere (Ref.~\onlinecite{Yogesh}). The obtained temperature dependence of the susceptibility is shown in Fig.~\ref{exc_Li213:order:sus}a). The high-temperature region (150 $<T<$ 370\ K) can be well described (red solid line) by a Curie-Weiss form $\chi=\chi_{0}+C/(T+\Theta_{\rm CW})$, with a fixed temperature-independent contribution $\chi_{0}=4.286\times10^{-4}$ emu~mol$^{-1}$~Oe$^{-1}$, $C=0.465$ emu~K~mol$^{-1}$~Oe$^{-1}$, and Curie-Weiss temperature $\Theta_{\rm CW}=-33.8$ K, consistent with previous reports (Ref.~\onlinecite{Yogesh}). The extracted effective magnetic moment $\mu_{\rm eff}=1.93$~$\mu_{\rm B}$ is close to the value 1.73~$\mu_{\rm B}$ expected for $J_{\rm eff}=1/2$ magnetic moments with g-factor $g=2$. Fig.~\ref{exc_Li213:order:sus}b) focuses on the
low-temperature behavior where a clear drop in susceptibility is observed near $T_{\rm N}=15$~K, characteristic of the onset of long-range magnetic order with antiferromagnetic correlations. We will show later that below this temperature, muon-spin relaxation measurements show evidence of static local magnetic fields. The data also showed a small hump near 5~K. Its origin is yet unclear and may indicate some small changes in the magnetic structure below this temperature. An anomaly near this temperature is also present in the $\mu$SR data that are to be discussed in the next Section.

%%%%%%%%%%%%%%%%%%%%%%%%%%%%%%%%%%%%%%%%%%%%%%%%%%%%%%%%%%%%%%%%%%%%
%%%%%%%%%%%%%%%%%%%%%%%%%%%%%%%%%%%%%%%%%%%%%%%%%%%%%%%%%%%%%%%%%%%%%

\section{Muon spin relaxation}
\label{sec:muSR}To further characterize the magnetic order zero-field muon-spin relaxation (ZF $\mu^{+}$SR) measurements (Ref.~\onlinecite{SB_muSR}) were performed on powder samples from the same batch using the GPS instrument at the Swiss Muon Source (S$\mu$S), Paul Scherrer Institut, Villigen, Switzerland. In a $\mu^{+}$SR experiment (Ref.~\onlinecite{SB_muSR}) spin polarized muons are implanted into the sample. The quantity of interest is the asymmetry $A(t)$, which is proportional to the spin polarization of the muon ensemble.

%%%%%%%%%%%%%%%%%%%%%%%%%%%%%%%%%%%%%%%%%%%%%%%%%%%%%%%%%%%%%%%%%%%
\begin{figure}[tbh]
\begin{center}
\includegraphics[width=\linewidth]{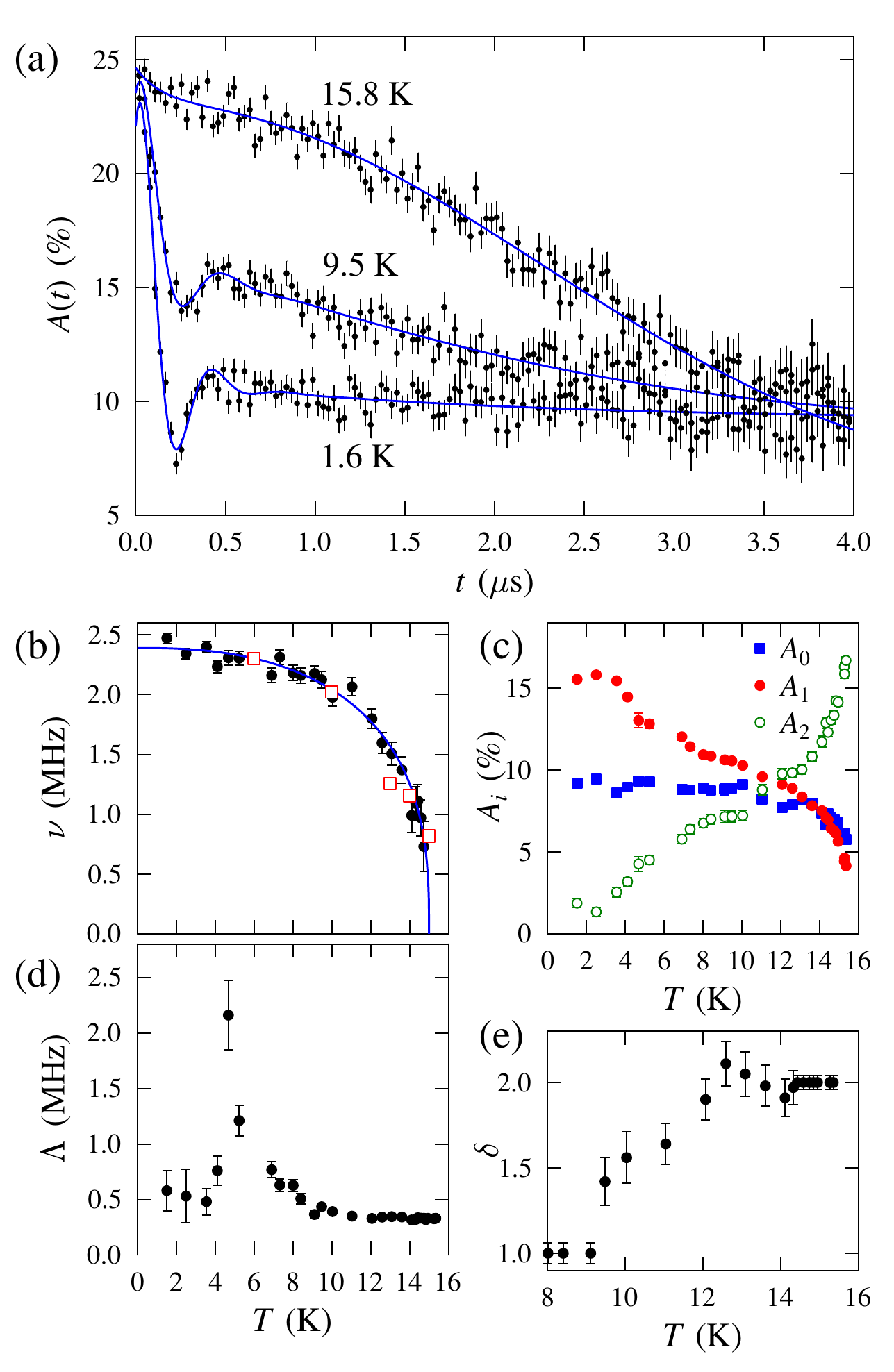}
\end{center}
%\vspace{0.5cm}
\caption {(color online) Muon spin relaxation results on $\alpha$-Li$_{2}$IrO$_{3}$ powder. a) Representative zero-field $\mu^{+}$SR spectra measured at several temperatures. Solid lines show fits to the functional form (\ref{eq:musr}) described in the text. b) Extracted temperature dependence of the muon precession frequency $\nu$ (filled circles) fitted to an order parameter behavior (solid line); open squares show the (scaled) order parameter extracted from the intensity of the magnetic Bragg peak observed in neutron powder diffraction data (squares from Fig. 4(inset) in Ref.~\onlinecite{Steph:Li213}). c) Temperature dependence of the fitted amplitudes $A_{0}$, $A_{1}$ and $A_{2}$. d) The relaxation rate $\Lambda$ and e) the stretching factor $\delta$.\label{exc_Li213:order:muSR}}
\end{figure}
%%%%%%%%%%%%%%%%%%%%%%%%%%%%%%%%%%%%%%%%%%%%%%%%%%%%%%%%%%%%%%%%%%%

Fig.~\ref{exc_Li213:order:muSR}a) shows representative ZF $\mu^{+}$SR spectra. Below $T_{\rm N}=15$\ K, a heavily damped oscillation in the muon asymmetry with a single frequency $\nu$ was found, characteristic of long range magnetic order. The measured data could be fitted for all $T < 15$~K with the function
\begin{equation}
A(t)=A_{0}+A_{1}e^{-\lambda t}\cos(2\pi\nu t+\phi)+A_{2}e^{-(\Lambda t)^{\delta}}.
\label{eq:musr}
\end{equation}
Here the first term, $A_{0}$, is a non-relaxing component accounting for a small fraction of muons that stop in the sample mount along with those muons whose spin components lie along the direction of the quasi-static local magnetic field. The second term $A_{1}$ is the oscillating component and the last term $A_{2}$ is a purely relaxing component which becomes more prominent upon increasing temperature at the expense of the oscillating component. A nonzero phase $\phi=-\pi/4$ was found to best fit the data. It is notable that this value is often indicative of an incommensurate magnetic ordering (Ref.~\onlinecite{Sugiyama:muSR:2004}) in agreement with resonant x-ray and neutron diffraction measurements, which observe the onset below $T_{\mathrm{N}}$ of a moment rotating structure with an incommensurate propagation vector (Ref.~\onlinecite{Steph:Li213}). For the fits, the total relaxing amplitude (at $t=0$) was kept fixed at 26.6~$\%$ and the relaxation rate of the oscillations was found to take a constant value of $\lambda=7.0$~MHz.

The temperature dependence of the extracted oscillation frequency is plotted in Fig.~\ref{exc_Li213:order:muSR}b) and is well described by the phenomenological function $\nu(T) = \nu(0) \left[1 - \left(T/T_{\mathrm{N}}\right)^{\alpha} \right]^{\beta}$, with $\nu(0)=2.39(2)$\ MHz, $\alpha=2.5$, $\beta=0.35(2)$ and $T_{\rm{N}}=15.0(1)$~K.  The overall temperature-dependence is consistent with the magnetic order
parameter extracted from neutron diffraction (open squares in
Fig.~\ref{exc_Li213:order:muSR}b), corresponding to $\sqrt{I}$, where
$I$ is the magnetic Bragg peak intensity
(Ref.~\onlinecite{Steph:Li213}). The onset temperature is consistent
with the location of the sharp anomaly observed in the susceptibility
data in Fig.~\ref{exc_Li213:order:sus}b). The amplitudes of the
components that reflect long-range magnetic order ($A_{0}$ and
$A_{1}$) are generally seen to decrease across the temperature range,
with the purely relaxing component $A_{2}$ increasing in their place
[see Fig.~\ref{exc_Li213:order:muSR}c)], which is indicative of
magnetically disordered regions increasing in volume within the sample
as the temperature rises. At the lowest temperature, the data confirm
long-range order throughout the sample. The value of $\nu(0)$ is
similar to that of the dominant precession frequency recently measured
in the $\beta$-phase of this compound\cite{Majumder2018}. We note that the relaxation rate $\Lambda$ of the purely relaxing component appears to have a maximum near 4.7~K [see Fig.~\ref{exc_Li213:order:muSR}d)], which coincides with the presence of a small anomaly in the susceptibility data [see Fig.~\ref{exc_Li213:order:sus}b)].

%%%%%%%%%%%%%%%%%%%%%%%%%%%%%%%%%%%%%%%%%%%%%%%%%%%%%
\begin{figure}[t!]
\begin{center}
\includegraphics[width=\linewidth]{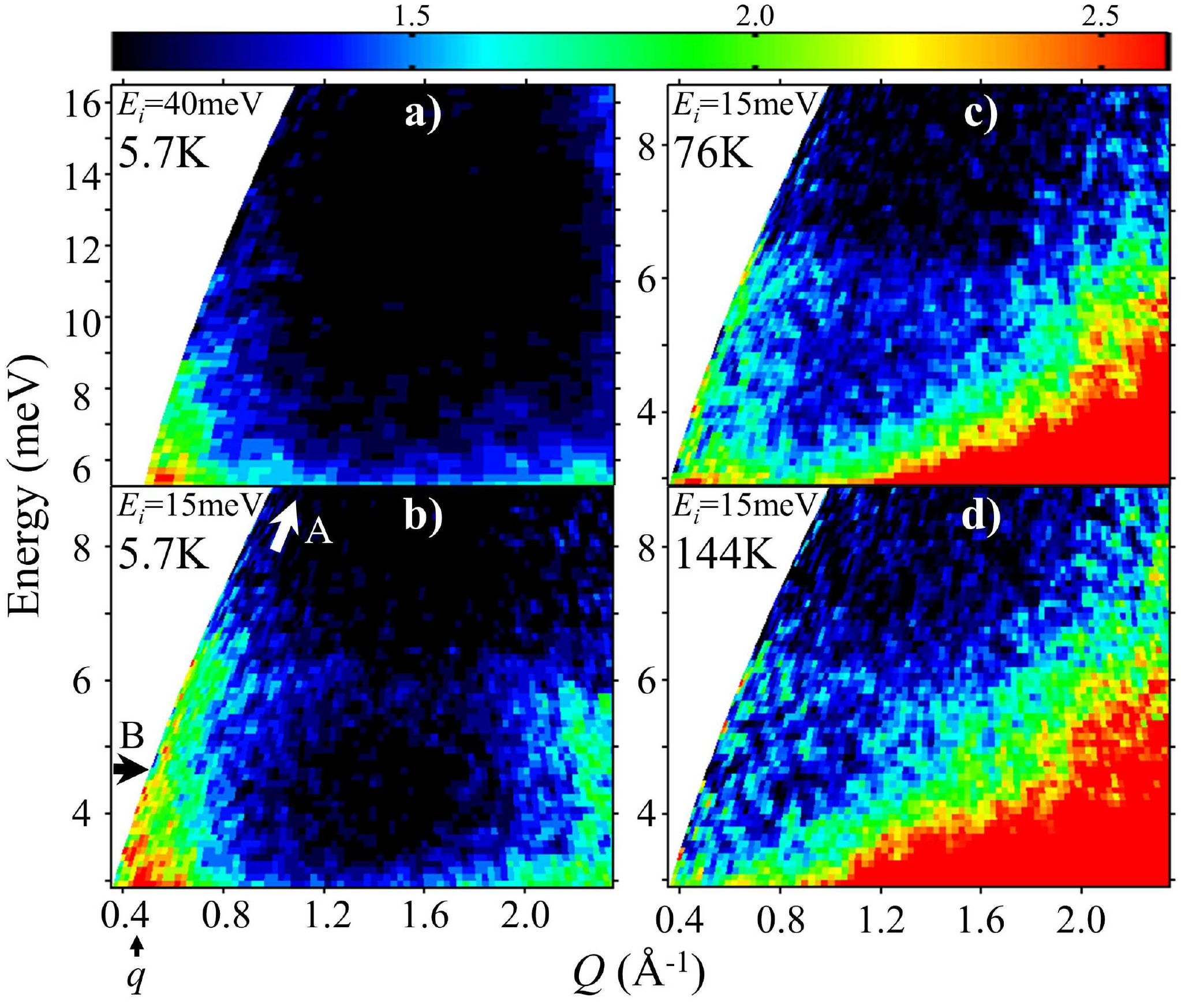}
\end{center}
\caption {(color online) Powder inelastic neutron scattering intensity as a function of wavevector and energy transfer. A dispersive inelastic signal centred near the magnetic Bragg peak wavevector $q$ (thick vertical arrow below the lower left corner) is clearly observed at 5.7~K deep in the magnetically ordered phase (panels a-b), and becomes damped out at high temperatures (panels c-d), as expected for a magnetic inelastic signal. The thick arrows labelled A and B in b) show directions along which the measured intensities are plotted in Fig.~\ref{exc:Li213:dispersion:cut}A-B. Data was collected with $E_i=40$~meV in a) and 15~meV in b-d). An overall scale factor was applied to the intensities in a) to match those in b) in the overlapping region.}
\label{exc:Li213:dispersion:15meV}
\end{figure}
%%%%%%%%%%%%%%%%%%%%%%%%%%%%%%%%%%%%%%%%%%%%%%%%%%%%%%%%%%%%%%%%%%%

%%%%%%%%%%%%%%%%%%%%%%%%%%%%%%%%%%%%%%%%%%%%%%%%%%%%%%%%%%%%%%%%%%%
\begin{figure}[t!]
\begin{center}
\includegraphics[width=\linewidth]{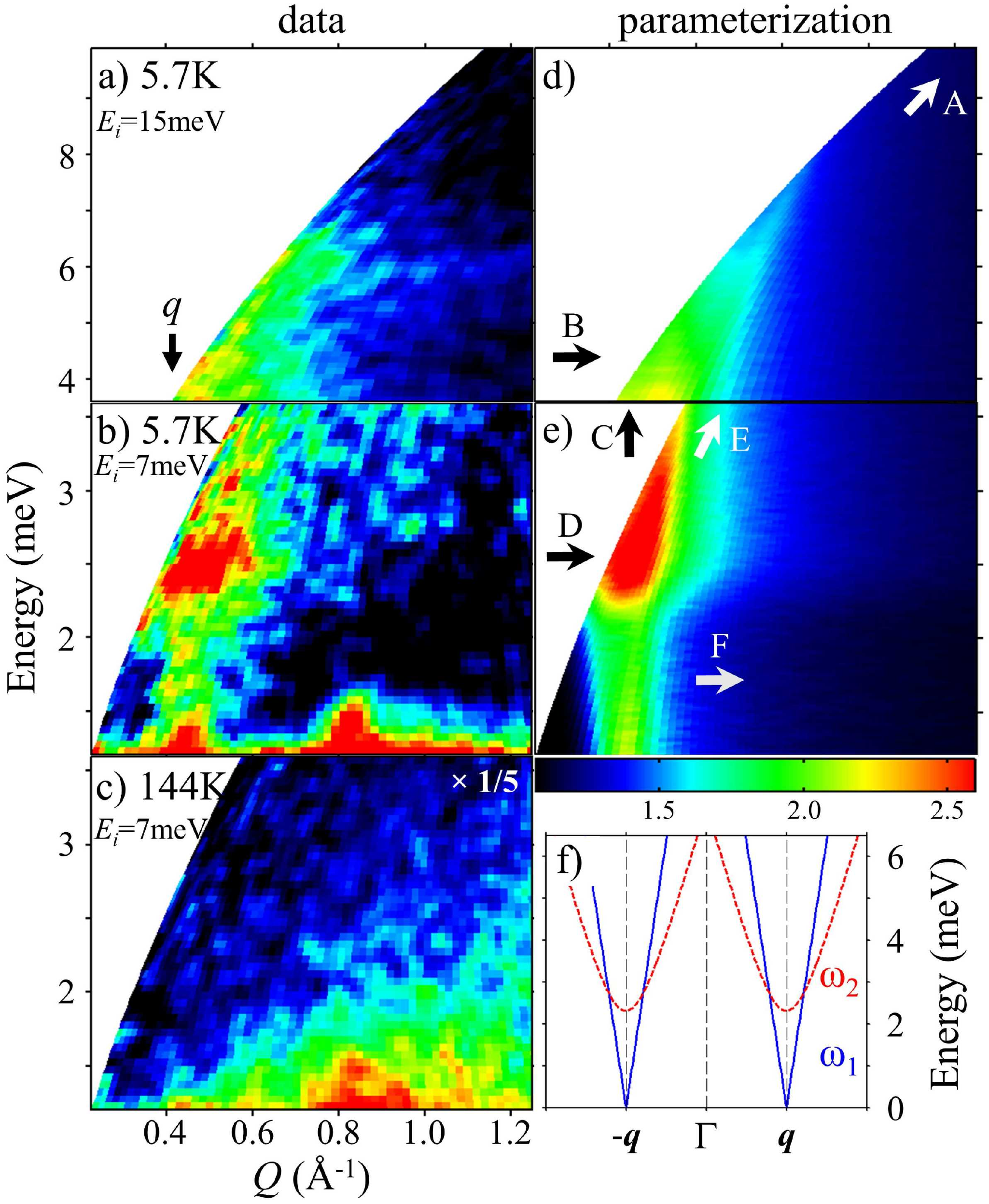}
\end{center}
\caption {(color online) a-b) Higher resolution measurements of the low-energy inelastic neutron scattering intensity. Near the magnetic ordering wavevector $q$ down (thick vertical arrow in a) an inelastic signal is observed down to the lowest accessible energy $\simeq$1~meV with a clear intensity increase near 2.3~meV; all this structure disappears upon heating to high temperatures (panel c), verifying its magnetic character. Intensities in c) have been scaled by a factor $1/5$ to make them comparable to the other panels. Data in a) and b) were collected under different instrumental conditions and for ease of visualization are shown as if they had a continuous energy axis between them with the intensities in the top panel multiplied by a single overall scale factor to best match those in the lower panel in the overlapping region. d-e) Calculated spherically-averaged magnetic inelastic scattering intensity (to be compared with the data in a-b) for the model cross-section discussed in the text with two modes with dispersions plotted in f). Thick arrows labelled A-F in d-e) show directions along which the measured and calculated intensity are plotted in Fig.~\ref{exc:Li213:dispersion:cut}A-F.} \label{exc:Li213:dispersion:data}
\end{figure}
%%%%%%%%%%%%%%%%%%%%%%%%%%%%%%%%%%%%%%%%%%%%%%%%%%%%%%%%%%%%%%%%%%%
%%%%%%%%%%%%%%%%%%%%%%%%%%%%%%%%%%%%%%%%%%%%%%%%%%%%%%%%%%%%%%%%%%%
\begin{figure}[t!]
\begin{center}
\includegraphics[width=\linewidth]{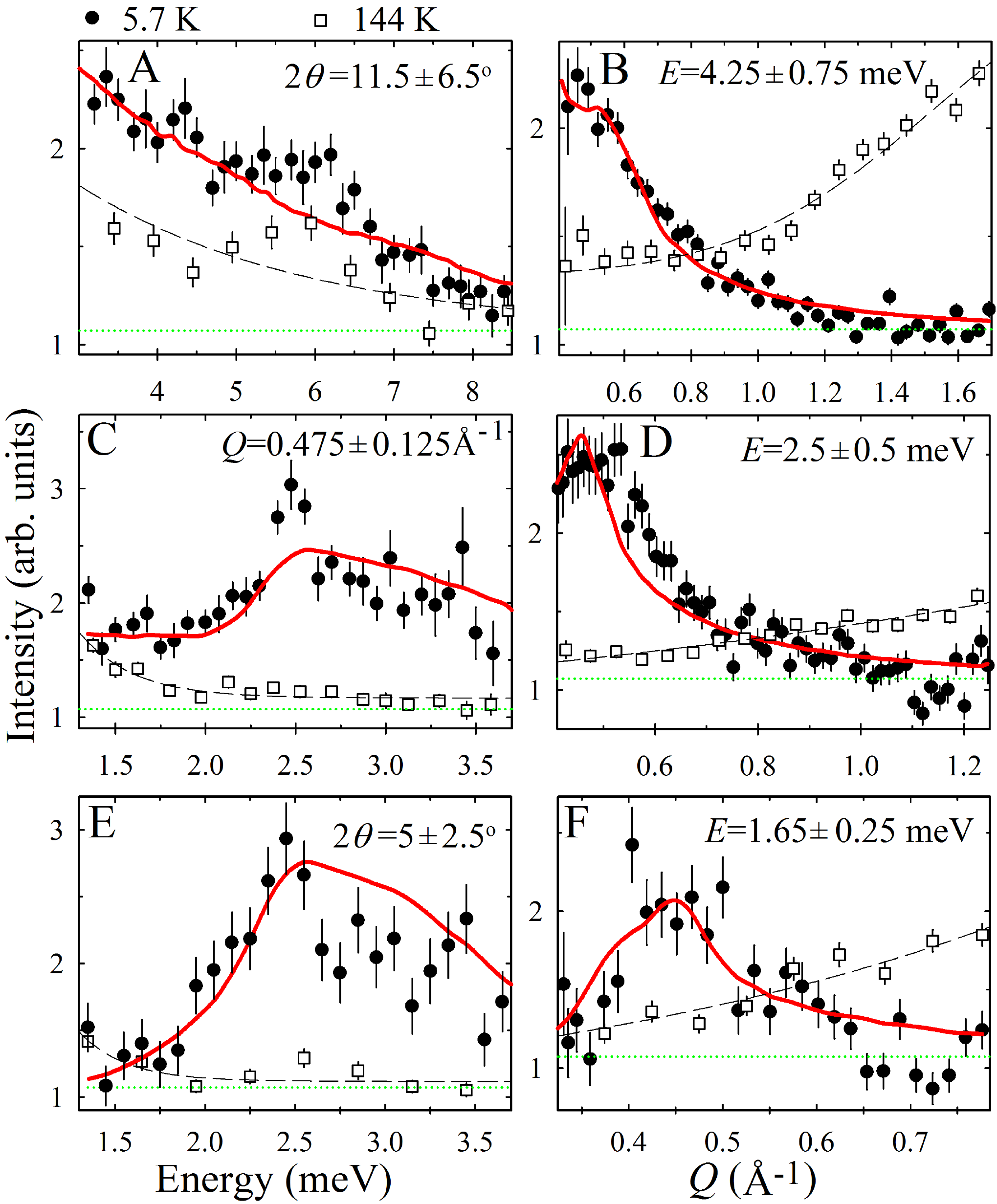}
\end{center}
\caption {(color online) Scans through the inelastic neutron scattering data in Figs.~\ref{exc:Li213:dispersion:15meV}b) and ~\ref{exc:Li213:dispersion:data}a-c) along directions indicated by the thick arrows labelled A-F in Fig.~\ref{exc:Li213:dispersion:data}d,e). Filled (open) symbols are data at 5.7~K (144~K) and solid red lines show the corresponding intensities in the empirical model parametrization defined in eqs.~(\ref{eq:disp}-\ref{eq:sqw}). Dashed lines are guides to the eye to indicate the trends in the 144~K data. Green dotted lines show the estimated non-magnetic background at low temperatures obtained by extrapolation from regions of high $Q$ or high $E$. Panel legends give the intensity integration ranges in wavevector $Q$, energy $E$ or total
scattering angle $2\theta$.} \label{exc:Li213:dispersion:cut}
\end{figure}
%%%%%%%%%%%%%%%%%%%%%%%%%%%%%%%%%%%%%%%%%%%%%%%%%%%%%%%%%%%%%%%%%%%

\section{Powder Inelastic Neutron Scattering}
\label{sec:INS} To probe the spin dynamics, inelastic neutron scattering measurements were performed using the high-flux, time-of-fight, direct-geometry chopper spectrometer MERLIN at ISIS. A fine powder of $\alpha$-Li$_2$IrO$_3$ (8.9~g) was placed in an annular can to minimize the strong neutron absorption from both Ir and Li (absorption cross-sections of 425 and 70.5~barns, respectively, for thermal neutrons). Cooling was provided by a closed cycle refrigerator with a base temperature of 5.7~K, well below $T_N$. The inelastic scattering was measured for incident neutrons of energies $E_i$~=~7, 15, 40 and 80~meV, and a clear inelastic magnetic signal was detected at low wavevector transfers for energies extending up to 12~meV. Most data were therefore collected with $E_i$~=~7 and 15~meV for which the instrumental energy resolution on the elastic line (FWHM) was $0.58(1)$ and $1.08(1)$~meV, respectively. Counting times per setting ranged between 17-25~hours at an average proton current of 150~$\mu$A.

Fig.~\ref{exc:Li213:dispersion:15meV} shows the wavevector ($Q$) and energy ($E$) dependence of the inelastic scattering at several temperatures. At low temperatures (panel b) a clear inelastic signal is visible at low wavevectors extending up in energy to at least 7~meV and centered at low energies near the magnitude $q$ (vertical arrow under the figure) of the incommensurate magnetic ordering wavevector $\bm{q}=(0.32(1),0,0)$ (Ref.~\onlinecite{Steph:Li213}). The scattering intensity in this region decreases upon increasing temperature and is completely damped out deep in the paramagnetic regime at 144~K (panel d), confirming its magnetic origin. In contrast, the intensity at large wavevectors and low energies strongly increases upon increasing temperature as expected for scattering processes involving phonons. Measurements with a higher incident neutron energy $E_i=40$~meV confirmed that the magnetic inelastic signal at low $Q$ extends in energy up to at least 12~meV [see Fig.~\ref{exc:Li213:dispersion:15meV}a)]. Higher resolution measurements (collected with $E_i=7$~meV), focusing on the low energy part of the spectrum, are shown in Fig.~\ref{exc:Li213:dispersion:data}b). Note the inelastic signal at low wavevectors near the magnetic ordering wavevector $q$ (thick vertical arrow above the data) with strong intensity near 2.3~meV and with a clear signal extending below this region, down to the lowest energies probed. All this structured inelastic signal becomes damped out upon heating to 144~K, Fig.~\ref{exc:Li213:dispersion:data}c), confirming its magnetic character.

\section{Parametrization of the low-energy spin dynamics by an
empirical spin-wave model}
\label{sec:par}

The low-energy inelastic magnetic response shows weak scattering intensity appearing to emerge out of the magnetic Bragg peak wavevector and extending up in energy, followed by an onset of much stronger scattering intensity above a gap. Those features resembles the generic structure of the low-energy spin excitations near the magnetic ordering wavevector in spiral-ordered magnets with easy-plane anisotropy. In that case the low-energy excitations near the ordering wavevector $\bm{q}$ contain a gapless (Goldstone) mode with a linear dispersion associated with long-wavelength spin oscillations confined to the spiral plane, and a gapped mode associated with fluctuations normal to the spiral plane. Inspired by this generic resemblance we empirically parameterize the low-energy inelastic data in terms of a minimal model with two dispersive modes (gapless $\hbar\omega_1$ and gapped $\hbar\omega_2$). For simplicity we consider both modes dispersing (isotropically) in the reciprocal $\bm{a}^{*}\bm{b}^{*}$ plane (as expected for magnetically decoupled honeycomb layers in the $ab$ plane), specifically
\begin{eqnarray}
%\hbar\omega(\bm{Q}) & = & \sqrt{v^2 |\bm{Q} - \bm{q}|^2+\Delta^2 \sin^2(\pi l)}, \nonumber \\
\hbar\omega_1(\bm{Q}) & = & v_1 |\bm{Q}_{\perp} - \bm{q}|, \nonumber \\
\hbar\omega_2(\bm{Q}) & = & \sqrt{v_2^2 \left|\bm{Q}_{\perp} -
\bm{q}\right|^{2}+\Delta^2}, \label{eq:disp}
\end{eqnarray}
where $\bm{q}$ is the incommensurate magnetic ordering wavevector (contained in the $\bm{a}^{*}\bm{b}^{*}$ plane), $\bm{Q}_{\perp}$ is the projection of the 3D wavevector $\bm{Q}$ onto the $\bm{a}^{*}\bm{b}^{*}$ plane, $v_{1,2}$ are the velocities of the two modes and $\Delta$ is the gap of the second mode. The above definition is for the case when $\bm{Q}_{\perp}$ is in the vicinity of $\bm{q}$. By symmetry, for wavevectors with $\bm{Q}_{\perp}$ in the vicinity of $-\bm{q}$, the same definition (\ref{eq:disp}) applies, but with $\bm{q}$ replaced by $-\bm{q}$. For this parametrization, the dispersions along the $-\bm{q}\rightarrow\Gamma\rightarrow\bm{q}$ direction are illustrated in Fig.~\ref{exc:Li213:dispersion:data}f). As a minimal model for the neutron scattering cross-section we assume an inverse energy intensity dependence (as generic for low-energy antiferromagnetic spin waves), allow for independent intensity pre-factors $A_{1,2}$ for the two modes, and we also assume that any effects of the intensity polarization dependence of the modes can be captured in a first approximation by a re-scaling of the intensity pre-factors $A_{1,2}$. Specifically, we assume the intensity dependence
\begin{eqnarray}
I({\bm Q},E)  = \left(\frac{g}{2} f(Q)\right)^2 & & \left[\frac{A_1}{E} ~ G(E-\hbar\omega_{1}(\bm{Q}))\right. \nonumber \\
& & + \left. \frac{A_2}{E} ~ G(E-\hbar\omega_{2}(\bm{Q}))\right], \label{eq:sqw}
\end{eqnarray}
where $f(Q)$ is the Ir$^{4+}$ spherical magnetic form factor (Ref.~\onlinecite{S_Irmff}) the $g$-factor was assumed to be equal to 2 and
$G(E)$ is a Gaussian function that reflects the finite instrumental energy resolution.

For comparison with the INS data at a given ($Q,E$) point the above equation was numerically averaged for a spherical distribution of wavevector transfers $\bm{Q}$ of fixed magnitude $Q$. The above parameterization could capture well the observed intensity dependence of the magnetic scattering in wavevector and energy. Representative values for the model parameters are $A_1/A_2=2.7$, $v_1=12.2$~meV\AA, $v_2=6.5$~meV\AA~ and gap $\Delta=2.3$~meV. The level of agreement obtained in this case can be seen by comparing the data in Fig.~\ref{exc:Li213:dispersion:data}a-b) with the calculation in panels d-e). The intensities in representative scans along energy and momentum directions are shown in Fig.~\ref{exc:Li213:dispersion:cut}A-F; the trends in the data (filled symbols) are well captured by the empirical parametrization (solid red lines). The constant-energy scan F below the energy gap $\Delta$ shows a clear peak in intensity near 0.45~\AA$^{-1}$ (which disappears at high temperatures - open squares) this intensity is associated with scattering from the $\hbar\omega_1$ mode. The energy scans C and E are directly sensitive to the gap $\Delta$ where a clear increase in scattering intensity is observed. In spite of its simplified nature, the empirical model with two modes with two-dimensional dispersions provides a good description of the general features of the magnetic inelastic scattering data from the lowest measured energies $\simeq$~1~meV up to intermediate energies $\simeq$~5~meV. We also compared the data with a modified model with isotropic three-dimensional dispersions for both modes (with $\bm{Q}_{\perp}$ replaced by $\bm{Q}$ in (\ref{eq:disp})), but this gave a worse fit to the experimental data, suggesting that a model with predominantly two-dimensional dispersions (as expected for nearly magnetically-decoupled layers) is a more suitable description.

%%%%%%%%%%%%%%%%%%%%%%%%%%%%%%%%%%%%%%%%%%%%%%%%%%%%%%%%%%%%%%%%%%%
\section{Pulsed-field magnetization}
\label{sec:mag}
%%%%%%%%%%%%%%%%%%%%%%%%%%%%%%%%%%%%%%%%%%%%%%%%%%%%%%%%%%%%%%%%%%%
\begin{figure}[t]
\begin{center}
\includegraphics[width=\linewidth] {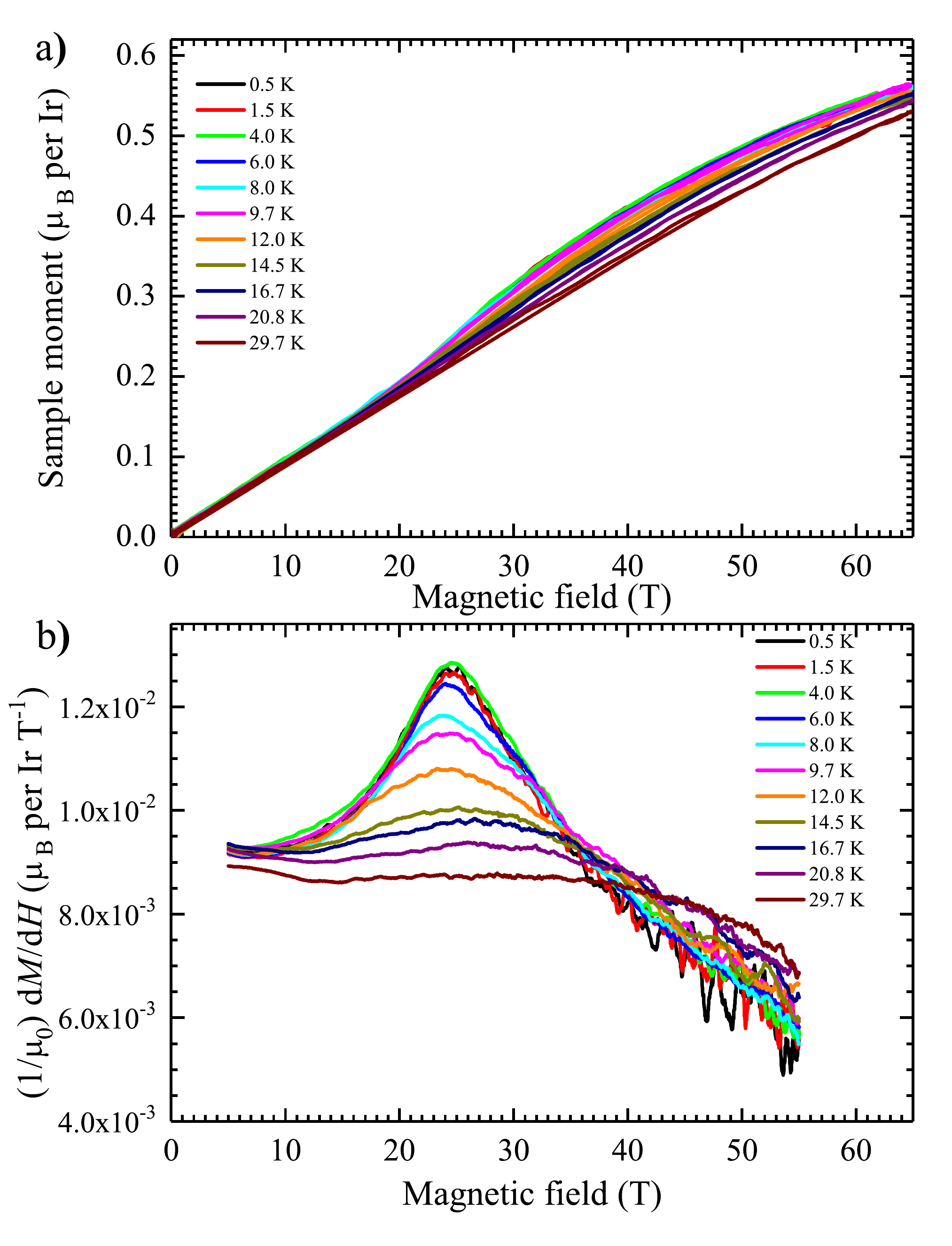}
\end{center}\caption {(color online) a) Magnetization as a function of
applied field $\mu_0H$ for a selection of temperatures. The curves are composites of pulsed-field data ($8-64$~T, averaged over multiple shots)
and low-field VSM data ($0-13$~T). %The thin lines are straight-line extrapolations of the zero-field susceptibility.
b) Differentials of the smoothed data sets in a).}
\label{fig:exp:mag}
\end{figure}
%%%%%%%%%%%%%%%%%%%%%%%%%%%%%%%%%%%%%%%%%%%%%%%%%%%%%%%%%%%%%%%%%%%

Pulsed field magnetization experiments were performed using the extraction magnetometer described in Ref.~\onlinecite{Goddard2008}, placed within a $^3$He cryostat with a base temperature of 0.4 K and the 65~T short-pulse magnet at NHMFL Los Alamos (Ref.~\onlinecite{Jaime2006}). The magnetization values measured in the pulsed-field experiments were calibrated into absolute units using data collected using a vibrating sample magnetometer (VSM). Magnetic field $M(H,T)$ data are shown for field sweeps up to 64~T at various constant temperatures $T$ in Fig.~\ref{fig:exp:mag}a). The low-temperature $M(H)$ curves show a pronounced steepening near a critical field $\mu_0H_\mathrm{C}\simeq25$~T, as characteristic of a field-induced phase transition. Differential susceptibility curves plotted in Fig.~\ref{fig:exp:mag}b) observe a well-defined ``peak'' at this field at low temperatures, with the peak decreasing in amplitude and broadening upon increasing temperature, with dramatic broadening above $\simeq$14.5~K. The differential susceptibility has a clear downwards trend upon increasing field much above the peak. At these high fields, the susceptibility is significantly suppressed compared to the value in the low-field spiral ordered region, but the absolute magnetization value is still only about half the expected saturation, assuming $g$-factor 2. Using the same $g$-factor the Zeeman energy of the critical field is $\simeq$$~2.9$~meV, comparable to the energy of the gapped mode in zero field ($\Delta\simeq2.3$~meV), suggesting that the mechanism of the phase transition could be related to the applied field suppressing this energy gap.

\section{Discussion}
\label{sec:disc}
It is interesting to set in context the magnetic phase transition observed near 25~T in $\alpha$-Li$_2$IrO$_3$. The behaviour of incommensurate spiral ordered magnets in magnetic field has been mostly investigated experimentally and theoretically (via mean-field analysis) for systems with frustrated isotropic, Heisenberg-type, interactions and some weak easy-plane anisotropy (Ref.~\onlinecite{Nagamiya}) In this case for fields applied normal to the rotation plane the moments cant towards the field direction to form a cone, typically stable up to the transition to magnetic saturation. For fields applied in the plane of rotation a succession of distinct phases is expected upon increasing field: spiral $\rightarrow$ cone $\rightarrow$ fan $\rightarrow$ saturated paramagnet; the first transition occurs at a field that overcomes a relatively small in-plane anisotropy energy, when the plane of moment rotation ``flops'' to be normal to the field axis and moments cant towards the field direction to form a cone. Those transitions would be detectable via anomalies in the magnetization curve as a function of field and for a powder sample one would expect to observe the spherical average of the magnetization curve for all possible directions of the applied field.

Interpreting the measurements on $\alpha$-Li$_2$IrO$_3$ in the above scenario, the transition at 25~T would be identified with the flopping of the moment rotation plane to become normal to the field axis with moments canted to form a cone phase; this transition would occur in the powder grains oriented to have the spin spiral almost parallel to the field direction with the grains oriented with the spiral plane near-normal to the field expected to have a smooth behaviour.

However, one difficulty with the above interpretation is that the susceptibility is seen experimentally to significantly decrease upon increasing field above the transition, see Fig.~\ref{fig:exp:mag}b), whereas a near-constant susceptibility would be expected at the mean field level throughout the cone phase (almost independent of the field orientation). Also, the type of  spiral order observed in $\alpha$-Li$_2$IrO$_3$ with counterrotating moments cannot be stabilized by Heisenberg-type interactions. In contrast, dominant (ferromagnetic) strongly-anisotropic Kitaev terms are required (Ref.~\onlinecite{Steph:Li213}) and in such cases the magnetic phase diagram in applied field could be qualitatively different, as suggested by recent theoretical proposals for three-dimensional hyperhoneycomb Kitaev magnets (Refs.~\onlinecite{Perkins2017, Perkins2018}). Experimental studies of other Kitaev materials such as $\alpha$-RuCl$_3$ (Refs.~\onlinecite{Johnson2015,Sears2017}), $\beta$-(Refs.~\onlinecite{Takayama:beta:Li213,Ruiz2017}) and potentially also $\gamma$-Li$_2$IrO$_3$ (Ref.~\onlinecite{Modic2017}), showed that relatively modest fields compared to the overall strength of the magnetic exchanges can suppress the spontaneous magnetic order altogether and stabilize instead an extended quantum paramagnetic region with significant quantum fluctuations. We hope our experimental studies will stimulate further theoretical studies of the magnetic phase diagram of honeycomb Kitaev systems with incommensurate counterrotating order, which may bring insight into the physics of the field-induced phase transition in $\alpha$-Li$_2$IrO$_3$, the magnetic properties above the critical field, and the potential role played by the low-energy gapped mode observed in the inelastic neutron scattering data in the mechanism of the field-induced transition. On the experimental side, future high-field magnetometry studies on single crystal samples, which have recently become available (Ref.~\onlinecite{Freund}) could provide important information on the magnetic properties above the critical transition field.

\section{Conclusions}
\label{sec:sum} To summarize, we have probed the spin dynamics in the layered honeycomb magnet $\alpha$-Li$_2$IrO$_3$, which displays an incommensurate magnetic structure with counterrotating moments, proposed to be stabilized by frustrated Kitaev interactions. The low-energy magnetic excitation spectrum observed by inelastic neutron scattering could be well parameterized by a gapless mode centered at the magnetic
ordering wavevector, associated with Goldstone mode fluctuations of the incommensurate order, and an additional intense gapped mode. In magnetization measurements in pulsed field we have observed evidence for a field-induced transition near 25~T to another magnetic phase with suppressed magnetic susceptibility and magnetization still well below the expected saturation.

\section{Acknowledgements}
\label{ack} We thank Paul Goddard and Itamar Kimchi for very helpful discussions. Work at Oxford was supported by ERC under Grant 788814 (EQFT) and by EPSRC (UK) under Grant Numbers EP/H014934/1, EP/M020517/1 and EP/N023803/1; at Durham under EP/G003092/1 and EP/N024028/1; and at ISIS by STFC (UK). We are grateful to Alex Amato and Hubertus Luetkens at S$\mu$S for experimental assistance. Work at Augsburg was supported by the Helmholtz Virtual Institute 521 ``New states of matter and their excitations" and the German Science Foundation through TRR-80. Work at Los Alamos National Laboratory (LANL) was supported by the U.S. Department of Energy (DoE) Basic Energy Science Field Work Proposal ``Science in 100 T''. The NHMFL facility at LANL is funded by the National Science Foundation Cooperative Agreement No. DMR-1157490, the State of Florida, and the U. S. DoE. In accordance with the EPSRC policy framework on research data, access to the data will be made available from Ref.~\onlinecite{data_archive}.

\bibliography{aLi2IrO3_dynamics_refs}

\begin{thebibliography}{50}%
\makeatletter
\providecommand \@ifxundefined [1]{%
 \@ifx{#1\undefined}
}%
\providecommand \@ifnum [1]{%
 \ifnum #1\expandafter \@firstoftwo
 \else \expandafter \@secondoftwo
 \fi
}%
\providecommand \@ifx [1]{%
 \ifx #1\expandafter \@firstoftwo
 \else \expandafter \@secondoftwo
 \fi
}%
\providecommand \natexlab [1]{#1}%
\providecommand \enquote  [1]{``#1''}%
\providecommand \bibnamefont  [1]{#1}%
\providecommand \bibfnamefont [1]{#1}%
\providecommand \citenamefont [1]{#1}%
\providecommand \href@noop [0]{\@secondoftwo}%
\providecommand \href [0]{\begingroup \@sanitize@url \@href}%
\providecommand \@href[1]{\@@startlink{#1}\@@href}%
\providecommand \@@href[1]{\endgroup#1\@@endlink}%
\providecommand \@sanitize@url [0]{\catcode `\\12\catcode `\$12\catcode
  `\&12\catcode `\#12\catcode `\^12\catcode `\_12\catcode `\%12\relax}%
\providecommand \@@startlink[1]{}%
\providecommand \@@endlink[0]{}%
\providecommand \url  [0]{\begingroup\@sanitize@url \@url }%
\providecommand \@url [1]{\endgroup\@href {#1}{\urlprefix }}%
\providecommand \urlprefix  [0]{URL }%
\providecommand \Eprint [0]{\href }%
\providecommand \doibase [0]{http://dx.doi.org/}%
\providecommand \selectlanguage [0]{\@gobble}%
\providecommand \bibinfo  [0]{\@secondoftwo}%
\providecommand \bibfield  [0]{\@secondoftwo}%
\providecommand \translation [1]{[#1]}%
\providecommand \BibitemOpen [0]{}%
\providecommand \bibitemStop [0]{}%
\providecommand \bibitemNoStop [0]{.\EOS\space}%
\providecommand \EOS [0]{\spacefactor3000\relax}%
\providecommand \BibitemShut  [1]{\csname bibitem#1\endcsname}%
\let\auto@bib@innerbib\@empty
%</preamble>
\bibitem [{\citenamefont {Rau}\ \emph {et~al.}(2016)\citenamefont {Rau},
  \citenamefont {Lee},\ and\ \citenamefont {Kee}}]{Rau2016}%
  \BibitemOpen
  \bibfield  {author} {\bibinfo {author} {\bibfnamefont {J.~G.}\ \bibnamefont
  {Rau}}, \bibinfo {author} {\bibfnamefont {E.~K.-H.}\ \bibnamefont {Lee}}, \
  and\ \bibinfo {author} {\bibfnamefont {H.-Y.}\ \bibnamefont {Kee}},\ }\href
  {\doibase 10.1146/annurev-conmatphys-031115-011319} {\bibfield  {journal}
  {\bibinfo  {journal} {Annu. Rev. Condens. Matter Phys.}\ }\textbf {\bibinfo
  {volume} {7}},\ \bibinfo {pages} {195} (\bibinfo {year} {2016})}\BibitemShut
  {NoStop}%
\bibitem [{\citenamefont {Winter}\ \emph {et~al.}(2016)\citenamefont {Winter},
  \citenamefont {Li}, \citenamefont {Jeschke},\ and\ \citenamefont
  {Valent\'{\i}}}]{Winter2016}%
  \BibitemOpen
  \bibfield  {author} {\bibinfo {author} {\bibfnamefont {S.~M.}\ \bibnamefont
  {Winter}}, \bibinfo {author} {\bibfnamefont {Y.}~\bibnamefont {Li}}, \bibinfo
  {author} {\bibfnamefont {H.~O.}\ \bibnamefont {Jeschke}}, \ and\ \bibinfo
  {author} {\bibfnamefont {R.}~\bibnamefont {Valent\'{\i}}},\ }\href {\doibase
  10.1103/PhysRevB.93.214431} {\bibfield  {journal} {\bibinfo  {journal} {Phys.
  Rev. B}\ }\textbf {\bibinfo {volume} {93}},\ \bibinfo {pages} {214431}
  (\bibinfo {year} {2016})}\BibitemShut {NoStop}%
\bibitem [{\citenamefont {Hermanns}\ \emph {et~al.}(2018)\citenamefont
  {Hermanns}, \citenamefont {Kimchi},\ and\ \citenamefont
  {Knolle}}]{Hermanns2018}%
  \BibitemOpen
  \bibfield  {author} {\bibinfo {author} {\bibfnamefont {M.}~\bibnamefont
  {Hermanns}}, \bibinfo {author} {\bibfnamefont {I.}~\bibnamefont {Kimchi}}, \
  and\ \bibinfo {author} {\bibfnamefont {J.}~\bibnamefont {Knolle}},\
  }\href@noop {} {\bibfield  {journal} {\bibinfo  {journal} {Annu. Rev.
  Condens. Matter Phys.}\ }\textbf {\bibinfo {volume} {9}},\  (\bibinfo {year}
  {2018})}\BibitemShut {NoStop}%
\bibitem [{\citenamefont {Kitaev}(2006)}]{kitaev}%
  \BibitemOpen
  \bibfield  {author} {\bibinfo {author} {\bibfnamefont {A.}~\bibnamefont
  {Kitaev}},\ }\href@noop {} {\bibfield  {journal} {\bibinfo  {journal} {Ann.
  Phys. (N. Y.)}\ }\textbf {\bibinfo {volume} {321}},\ \bibinfo {pages} {2}
  (\bibinfo {year} {2006})}\BibitemShut {NoStop}%
\bibitem [{\citenamefont {Knolle}\ \emph {et~al.}(2014)\citenamefont {Knolle},
  \citenamefont {Kovrizhin}, \citenamefont {Chalker},\ and\ \citenamefont
  {Moessner}}]{knolle}%
  \BibitemOpen
  \bibfield  {author} {\bibinfo {author} {\bibfnamefont {J.}~\bibnamefont
  {Knolle}}, \bibinfo {author} {\bibfnamefont {D.~L.}\ \bibnamefont
  {Kovrizhin}}, \bibinfo {author} {\bibfnamefont {J.~T.}\ \bibnamefont
  {Chalker}}, \ and\ \bibinfo {author} {\bibfnamefont {R.}~\bibnamefont
  {Moessner}},\ }\href@noop {} {\bibfield  {journal} {\bibinfo  {journal}
  {Phys. Rev. Lett.}\ }\textbf {\bibinfo {volume} {112}},\ \bibinfo {pages}
  {207203} (\bibinfo {year} {2014})}\BibitemShut {NoStop}%
\bibitem [{\citenamefont {Chaloupka}\ \emph {et~al.}(2013)\citenamefont
  {Chaloupka}, \citenamefont {Jackeli},\ and\ \citenamefont
  {Khaliullin}}]{Jackeli:zz}%
  \BibitemOpen
  \bibfield  {author} {\bibinfo {author} {\bibfnamefont {J.}~\bibnamefont
  {Chaloupka}}, \bibinfo {author} {\bibfnamefont {G.}~\bibnamefont {Jackeli}},
  \ and\ \bibinfo {author} {\bibfnamefont {G.}~\bibnamefont {Khaliullin}},\
  }\href@noop {} {\bibfield  {journal} {\bibinfo  {journal} {Phys. Rev. Lett}\
  }\textbf {\bibinfo {volume} {110}},\ \bibinfo {pages} {097204} (\bibinfo
  {year} {2013})}\BibitemShut {NoStop}%
\bibitem [{\citenamefont {Singh}\ and\ \citenamefont
  {Gegenwart}(2010)}]{Gegenwart2010}%
  \BibitemOpen
  \bibfield  {author} {\bibinfo {author} {\bibfnamefont {Y.}~\bibnamefont
  {Singh}}\ and\ \bibinfo {author} {\bibfnamefont {P.}~\bibnamefont
  {Gegenwart}},\ }\href@noop {} {\bibfield  {journal} {\bibinfo  {journal}
  {Phys. Rev. B}\ }\textbf {\bibinfo {volume} {82}},\ \bibinfo {pages} {064412}
  (\bibinfo {year} {2010})}\BibitemShut {NoStop}%
\bibitem [{\citenamefont {Plumb}\ \emph {et~al.}(2014)\citenamefont {Plumb},
  \citenamefont {Clancy}, \citenamefont {Sandilands}, \citenamefont {Shankar},
  \citenamefont {Hu}, \citenamefont {Burch}, \citenamefont {Kee},\ and\
  \citenamefont {Kim}}]{Plumb2014}%
  \BibitemOpen
  \bibfield  {author} {\bibinfo {author} {\bibfnamefont {K.~W.}\ \bibnamefont
  {Plumb}}, \bibinfo {author} {\bibfnamefont {J.~P.}\ \bibnamefont {Clancy}},
  \bibinfo {author} {\bibfnamefont {L.~J.}\ \bibnamefont {Sandilands}},
  \bibinfo {author} {\bibfnamefont {V.~V.}\ \bibnamefont {Shankar}}, \bibinfo
  {author} {\bibfnamefont {Y.~F.}\ \bibnamefont {Hu}}, \bibinfo {author}
  {\bibfnamefont {K.~S.}\ \bibnamefont {Burch}}, \bibinfo {author}
  {\bibfnamefont {H.-Y.}\ \bibnamefont {Kee}}, \ and\ \bibinfo {author}
  {\bibfnamefont {Y.-J.}\ \bibnamefont {Kim}},\ }\href {\doibase
  10.1103/PhysRevB.90.041112} {\bibfield  {journal} {\bibinfo  {journal} {Phys.
  Rev. B}\ }\textbf {\bibinfo {volume} {90}},\ \bibinfo {pages} {041112}
  (\bibinfo {year} {2014})}\BibitemShut {NoStop}%
\bibitem [{\citenamefont {Singh}\ \emph {et~al.}(2012)\citenamefont {Singh},
  \citenamefont {Manni}, \citenamefont {Reuther}, \citenamefont {Berlijn},
  \citenamefont {Thomale}, \citenamefont {Ku}, \citenamefont {Trebst},\ and\
  \citenamefont {Gegenwart}}]{Yogesh}%
  \BibitemOpen
  \bibfield  {author} {\bibinfo {author} {\bibfnamefont {Y.}~\bibnamefont
  {Singh}}, \bibinfo {author} {\bibfnamefont {S.}~\bibnamefont {Manni}},
  \bibinfo {author} {\bibfnamefont {J.}~\bibnamefont {Reuther}}, \bibinfo
  {author} {\bibfnamefont {T.}~\bibnamefont {Berlijn}}, \bibinfo {author}
  {\bibfnamefont {R.}~\bibnamefont {Thomale}}, \bibinfo {author} {\bibfnamefont
  {W.}~\bibnamefont {Ku}}, \bibinfo {author} {\bibfnamefont {S.}~\bibnamefont
  {Trebst}}, \ and\ \bibinfo {author} {\bibfnamefont {P.}~\bibnamefont
  {Gegenwart}},\ }\href {\doibase 10.1103/PhysRevLett.108.127203} {\bibfield
  {journal} {\bibinfo  {journal} {Phys. Rev. Lett.}\ }\textbf {\bibinfo
  {volume} {108}},\ \bibinfo {pages} {127203} (\bibinfo {year}
  {2012})}\BibitemShut {NoStop}%
\bibitem [{\citenamefont {Takayama}\ \emph {et~al.}(2015)\citenamefont
  {Takayama}, \citenamefont {Kato}, \citenamefont {Dinnebier}, \citenamefont
  {Nuss}, \citenamefont {Kono}, \citenamefont {Veiga}, \citenamefont {Fabbris},
  \citenamefont {Haskel},\ and\ \citenamefont {Takagi}}]{Takayama:beta:Li213}%
  \BibitemOpen
  \bibfield  {author} {\bibinfo {author} {\bibfnamefont {T.}~\bibnamefont
  {Takayama}}, \bibinfo {author} {\bibfnamefont {A.}~\bibnamefont {Kato}},
  \bibinfo {author} {\bibfnamefont {R.}~\bibnamefont {Dinnebier}}, \bibinfo
  {author} {\bibfnamefont {J.}~\bibnamefont {Nuss}}, \bibinfo {author}
  {\bibfnamefont {H.}~\bibnamefont {Kono}}, \bibinfo {author} {\bibfnamefont
  {L.~S.~I.}\ \bibnamefont {Veiga}}, \bibinfo {author} {\bibfnamefont
  {G.}~\bibnamefont {Fabbris}}, \bibinfo {author} {\bibfnamefont
  {D.}~\bibnamefont {Haskel}}, \ and\ \bibinfo {author} {\bibfnamefont
  {H.}~\bibnamefont {Takagi}},\ }\href {\doibase
  10.1103/PhysRevLett.114.077202} {\bibfield  {journal} {\bibinfo  {journal}
  {Phys. Rev. Lett.}\ }\textbf {\bibinfo {volume} {114}},\ \bibinfo {pages}
  {077202} (\bibinfo {year} {2015})}\BibitemShut {NoStop}%
\bibitem [{\citenamefont {Modic}\ \emph {et~al.}(2014)\citenamefont {Modic},
  \citenamefont {Smidt}, \citenamefont {Kimchi}, \citenamefont {Breznay},
  \citenamefont {Biffin}, \citenamefont {Choi}, \citenamefont {Johnson},
  \citenamefont {Coldea}, \citenamefont {Watkins-Curry}, \citenamefont
  {McCandless}, \citenamefont {Chan}, \citenamefont {Gandara}, \citenamefont
  {Islam}, \citenamefont {Vishwanath}, \citenamefont {Shekhter}, \citenamefont
  {McDonald},\ and\ \citenamefont {Analytis}}]{modic}%
  \BibitemOpen
  \bibfield  {author} {\bibinfo {author} {\bibfnamefont {K.~A.}\ \bibnamefont
  {Modic}}, \bibinfo {author} {\bibfnamefont {T.~E.}\ \bibnamefont {Smidt}},
  \bibinfo {author} {\bibfnamefont {I.}~\bibnamefont {Kimchi}}, \bibinfo
  {author} {\bibfnamefont {N.~P.}\ \bibnamefont {Breznay}}, \bibinfo {author}
  {\bibfnamefont {A.}~\bibnamefont {Biffin}}, \bibinfo {author} {\bibfnamefont
  {S.}~\bibnamefont {Choi}}, \bibinfo {author} {\bibfnamefont {R.~D.}\
  \bibnamefont {Johnson}}, \bibinfo {author} {\bibfnamefont {R.}~\bibnamefont
  {Coldea}}, \bibinfo {author} {\bibfnamefont {P.}~\bibnamefont
  {Watkins-Curry}}, \bibinfo {author} {\bibfnamefont {G.~T.}\ \bibnamefont
  {McCandless}}, \bibinfo {author} {\bibfnamefont {J.~Y.}\ \bibnamefont
  {Chan}}, \bibinfo {author} {\bibfnamefont {F.}~\bibnamefont {Gandara}},
  \bibinfo {author} {\bibfnamefont {Z.}~\bibnamefont {Islam}}, \bibinfo
  {author} {\bibfnamefont {A.}~\bibnamefont {Vishwanath}}, \bibinfo {author}
  {\bibfnamefont {A.}~\bibnamefont {Shekhter}}, \bibinfo {author}
  {\bibfnamefont {R.~D.}\ \bibnamefont {McDonald}}, \ and\ \bibinfo {author}
  {\bibfnamefont {J.~G.}\ \bibnamefont {Analytis}},\ }\href@noop {} {\bibfield
  {journal} {\bibinfo  {journal} {Nat. Commun.}\ }\textbf {\bibinfo {volume}
  {5}},\ \bibinfo {pages} {4203} (\bibinfo {year} {2014})}\BibitemShut
  {NoStop}%
\bibitem [{\citenamefont {Liu}\ \emph {et~al.}(2011)\citenamefont {Liu},
  \citenamefont {Berlijn}, \citenamefont {Yin}, \citenamefont {Ku},
  \citenamefont {Tsvelik}, \citenamefont {Kim}, \citenamefont {Gretarsson},
  \citenamefont {Singh}, \citenamefont {Gegenwart},\ and\ \citenamefont
  {Hill}}]{Liu2011}%
  \BibitemOpen
  \bibfield  {author} {\bibinfo {author} {\bibfnamefont {X.}~\bibnamefont
  {Liu}}, \bibinfo {author} {\bibfnamefont {T.}~\bibnamefont {Berlijn}},
  \bibinfo {author} {\bibfnamefont {W.-G.}\ \bibnamefont {Yin}}, \bibinfo
  {author} {\bibfnamefont {W.}~\bibnamefont {Ku}}, \bibinfo {author}
  {\bibfnamefont {A.}~\bibnamefont {Tsvelik}}, \bibinfo {author} {\bibfnamefont
  {Y.-J.}\ \bibnamefont {Kim}}, \bibinfo {author} {\bibfnamefont
  {H.}~\bibnamefont {Gretarsson}}, \bibinfo {author} {\bibfnamefont
  {Y.}~\bibnamefont {Singh}}, \bibinfo {author} {\bibfnamefont
  {P.}~\bibnamefont {Gegenwart}}, \ and\ \bibinfo {author} {\bibfnamefont
  {J.~P.}\ \bibnamefont {Hill}},\ }\href {\doibase 10.1103/PhysRevB.83.220403}
  {\bibfield  {journal} {\bibinfo  {journal} {Phys. Rev. B}\ }\textbf {\bibinfo
  {volume} {83}},\ \bibinfo {pages} {220403} (\bibinfo {year}
  {2011})}\BibitemShut {NoStop}%
\bibitem [{\citenamefont {Choi}\ \emph {et~al.}(2012)\citenamefont {Choi},
  \citenamefont {Coldea}, \citenamefont {Kolmogorov}, \citenamefont
  {Lancaster}, \citenamefont {Mazin}, \citenamefont {Blundell}, \citenamefont
  {Radaelli}, \citenamefont {Singh}, \citenamefont {Gegenwart}, \citenamefont
  {Choi}, \citenamefont {Cheong}, \citenamefont {Baker}, \citenamefont
  {Stock},\ and\ \citenamefont {Taylor}}]{Na213_INS_2012}%
  \BibitemOpen
  \bibfield  {author} {\bibinfo {author} {\bibfnamefont {S.~K.}\ \bibnamefont
  {Choi}}, \bibinfo {author} {\bibfnamefont {R.}~\bibnamefont {Coldea}},
  \bibinfo {author} {\bibfnamefont {A.~N.}\ \bibnamefont {Kolmogorov}},
  \bibinfo {author} {\bibfnamefont {T.}~\bibnamefont {Lancaster}}, \bibinfo
  {author} {\bibfnamefont {I.~I.}\ \bibnamefont {Mazin}}, \bibinfo {author}
  {\bibfnamefont {S.~J.}\ \bibnamefont {Blundell}}, \bibinfo {author}
  {\bibfnamefont {P.~G.}\ \bibnamefont {Radaelli}}, \bibinfo {author}
  {\bibfnamefont {Y.}~\bibnamefont {Singh}}, \bibinfo {author} {\bibfnamefont
  {P.}~\bibnamefont {Gegenwart}}, \bibinfo {author} {\bibfnamefont {K.~R.}\
  \bibnamefont {Choi}}, \bibinfo {author} {\bibfnamefont {S.-W.}\ \bibnamefont
  {Cheong}}, \bibinfo {author} {\bibfnamefont {P.~J.}\ \bibnamefont {Baker}},
  \bibinfo {author} {\bibfnamefont {C.}~\bibnamefont {Stock}}, \ and\ \bibinfo
  {author} {\bibfnamefont {J.}~\bibnamefont {Taylor}},\ }\href@noop {}
  {\bibfield  {journal} {\bibinfo  {journal} {Phys. Rev. Lett.}\ }\textbf
  {\bibinfo {volume} {108}},\ \bibinfo {pages} {127204} (\bibinfo {year}
  {2012})}\BibitemShut {NoStop}%
\bibitem [{\citenamefont {Ye}\ \emph {et~al.}(2012)\citenamefont {Ye},
  \citenamefont {Chi}, \citenamefont {Cao}, \citenamefont {Chakoumakos},
  \citenamefont {Fernandez-Baca}, \citenamefont {Custelcean}, \citenamefont
  {Qi}, \citenamefont {Korneta},\ and\ \citenamefont {Cao}}]{Ye2012}%
  \BibitemOpen
  \bibfield  {author} {\bibinfo {author} {\bibfnamefont {F.}~\bibnamefont
  {Ye}}, \bibinfo {author} {\bibfnamefont {S.}~\bibnamefont {Chi}}, \bibinfo
  {author} {\bibfnamefont {H.}~\bibnamefont {Cao}}, \bibinfo {author}
  {\bibfnamefont {B.~C.}\ \bibnamefont {Chakoumakos}}, \bibinfo {author}
  {\bibfnamefont {J.~A.}\ \bibnamefont {Fernandez-Baca}}, \bibinfo {author}
  {\bibfnamefont {R.}~\bibnamefont {Custelcean}}, \bibinfo {author}
  {\bibfnamefont {T.~F.}\ \bibnamefont {Qi}}, \bibinfo {author} {\bibfnamefont
  {O.~B.}\ \bibnamefont {Korneta}}, \ and\ \bibinfo {author} {\bibfnamefont
  {G.}~\bibnamefont {Cao}},\ }\href {\doibase 10.1103/PhysRevB.85.180403}
  {\bibfield  {journal} {\bibinfo  {journal} {Phys. Rev. B}\ }\textbf {\bibinfo
  {volume} {85}},\ \bibinfo {pages} {180403} (\bibinfo {year}
  {2012})}\BibitemShut {NoStop}%
\bibitem [{\citenamefont {Hwan~Chun}\ \emph {et~al.}(2015)\citenamefont
  {Hwan~Chun}, \citenamefont {Kim}, \citenamefont {Kim}, \citenamefont {Zheng},
  \citenamefont {Stoumpos}, \citenamefont {Malliakas}, \citenamefont
  {Mitchell}, \citenamefont {Mehlawat}, \citenamefont {Singh}, \citenamefont
  {Choi}, \citenamefont {Gog}, \citenamefont {Al-Zein}, \citenamefont {Sala},
  \citenamefont {Krisch}, \citenamefont {Jackeli}, \citenamefont {Khaliullin},\
  and\ \citenamefont {Kim}}]{Hwan2015}%
  \BibitemOpen
  \bibfield  {author} {\bibinfo {author} {\bibfnamefont {S.}~\bibnamefont
  {Hwan~Chun}}, \bibinfo {author} {\bibfnamefont {J.-W.}\ \bibnamefont {Kim}},
  \bibinfo {author} {\bibfnamefont {J.}~\bibnamefont {Kim}}, \bibinfo {author}
  {\bibfnamefont {H.}~\bibnamefont {Zheng}}, \bibinfo {author} {\bibfnamefont
  {C.~C.}\ \bibnamefont {Stoumpos}}, \bibinfo {author} {\bibfnamefont {C.~D.}\
  \bibnamefont {Malliakas}}, \bibinfo {author} {\bibfnamefont {J.~F.}\
  \bibnamefont {Mitchell}}, \bibinfo {author} {\bibfnamefont {K.}~\bibnamefont
  {Mehlawat}}, \bibinfo {author} {\bibfnamefont {Y.}~\bibnamefont {Singh}},
  \bibinfo {author} {\bibfnamefont {Y.}~\bibnamefont {Choi}}, \bibinfo {author}
  {\bibfnamefont {T.}~\bibnamefont {Gog}}, \bibinfo {author} {\bibfnamefont
  {A.}~\bibnamefont {Al-Zein}}, \bibinfo {author} {\bibfnamefont {M.~M.}\
  \bibnamefont {Sala}}, \bibinfo {author} {\bibfnamefont {J.}~\bibnamefont
  {Krisch}, \bibfnamefont {M.~Chaloupka}}, \bibinfo {author} {\bibfnamefont
  {G.}~\bibnamefont {Jackeli}}, \bibinfo {author} {\bibfnamefont
  {G.}~\bibnamefont {Khaliullin}}, \ and\ \bibinfo {author} {\bibfnamefont
  {B.~J.}\ \bibnamefont {Kim}},\ }\href {\doibase 10.1038/nphys3322} {\bibfield
   {journal} {\bibinfo  {journal} {Nature Phys.}\ }\textbf {\bibinfo {volume}
  {11}},\ \bibinfo {pages} {462} (\bibinfo {year} {2015})}\BibitemShut
  {NoStop}%
\bibitem [{\citenamefont {Das}\ \emph {et~al.}()\citenamefont {Das},
  \citenamefont {Kundu}, \citenamefont {Zhu}, \citenamefont {Mun},
  \citenamefont {McDonald}, \citenamefont {Li}, \citenamefont {Balicas},
  \citenamefont {McCollam}, \citenamefont {Cao}, \citenamefont {Rau},
  \citenamefont {Kee}, \citenamefont {Tripathi},\ and\ \citenamefont
  {Sebastian}}]{Das2017}%
  \BibitemOpen
  \bibfield  {author} {\bibinfo {author} {\bibfnamefont {S.}~\bibnamefont
  {Das}}, \bibinfo {author} {\bibfnamefont {S.}~\bibnamefont {Kundu}}, \bibinfo
  {author} {\bibfnamefont {Z.}~\bibnamefont {Zhu}}, \bibinfo {author}
  {\bibfnamefont {E.}~\bibnamefont {Mun}}, \bibinfo {author} {\bibfnamefont
  {R.}~\bibnamefont {McDonald}}, \bibinfo {author} {\bibfnamefont
  {G.}~\bibnamefont {Li}}, \bibinfo {author} {\bibfnamefont {L.}~\bibnamefont
  {Balicas}}, \bibinfo {author} {\bibfnamefont {A.}~\bibnamefont {McCollam}},
  \bibinfo {author} {\bibfnamefont {G.}~\bibnamefont {Cao}}, \bibinfo {author}
  {\bibfnamefont {J.}~\bibnamefont {Rau}}, \bibinfo {author} {\bibfnamefont
  {H.-Y.}\ \bibnamefont {Kee}}, \bibinfo {author} {\bibfnamefont
  {V.}~\bibnamefont {Tripathi}}, \ and\ \bibinfo {author} {\bibfnamefont
  {S.}~\bibnamefont {Sebastian}},\ }\href@noop {} {\bibinfo  {journal}
  {arXiv:1708.03235}\ }\BibitemShut {NoStop}%
\bibitem [{\citenamefont {Johnson}\ \emph {et~al.}(2015)\citenamefont
  {Johnson}, \citenamefont {Williams}, \citenamefont {Haghighirad},
  \citenamefont {Singleton}, \citenamefont {Zapf}, \citenamefont {Manuel},
  \citenamefont {Mazin}, \citenamefont {Li}, \citenamefont {Jeschke},
  \citenamefont {Valent\'{\i}},\ and\ \citenamefont {Coldea}}]{Johnson2015}%
  \BibitemOpen
\bibfield  {journal} {  }\bibfield  {author} {\bibinfo {author} {\bibfnamefont
  {R.~D.}\ \bibnamefont {Johnson}}, \bibinfo {author} {\bibfnamefont {S.~C.}\
  \bibnamefont {Williams}}, \bibinfo {author} {\bibfnamefont {A.~A.}\
  \bibnamefont {Haghighirad}}, \bibinfo {author} {\bibfnamefont
  {J.}~\bibnamefont {Singleton}}, \bibinfo {author} {\bibfnamefont
  {V.}~\bibnamefont {Zapf}}, \bibinfo {author} {\bibfnamefont {P.}~\bibnamefont
  {Manuel}}, \bibinfo {author} {\bibfnamefont {I.~I.}\ \bibnamefont {Mazin}},
  \bibinfo {author} {\bibfnamefont {Y.}~\bibnamefont {Li}}, \bibinfo {author}
  {\bibfnamefont {H.~O.}\ \bibnamefont {Jeschke}}, \bibinfo {author}
  {\bibfnamefont {R.}~\bibnamefont {Valent\'{\i}}}, \ and\ \bibinfo {author}
  {\bibfnamefont {R.}~\bibnamefont {Coldea}},\ }\href {\doibase
  10.1103/PhysRevB.92.235119} {\bibfield  {journal} {\bibinfo  {journal} {Phys.
  Rev. B}\ }\textbf {\bibinfo {volume} {92}},\ \bibinfo {pages} {235119}
  (\bibinfo {year} {2015})}\BibitemShut {NoStop}%
\bibitem [{\citenamefont {Banerjee}\ \emph {et~al.}(2016)\citenamefont
  {Banerjee}, \citenamefont {Bridges}, \citenamefont {Yan}, \citenamefont
  {Aczel}, \citenamefont {Li}, \citenamefont {Stone}, \citenamefont {Granroth},
  \citenamefont {Lumsden}, \citenamefont {Yiu}, \citenamefont {Knolle},
  \citenamefont {Bhattacharjee}, \citenamefont {Kovrizhin}, \citenamefont
  {Moessner}, \citenamefont {Tennant}, \citenamefont {Mandrus},\ and\
  \citenamefont {Nagler}}]{Banerjee2016}%
  \BibitemOpen
  \bibfield  {author} {\bibinfo {author} {\bibfnamefont {A.}~\bibnamefont
  {Banerjee}}, \bibinfo {author} {\bibfnamefont {C.~A.}\ \bibnamefont
  {Bridges}}, \bibinfo {author} {\bibfnamefont {J.-Q.}\ \bibnamefont {Yan}},
  \bibinfo {author} {\bibfnamefont {A.~A.}\ \bibnamefont {Aczel}}, \bibinfo
  {author} {\bibfnamefont {L.}~\bibnamefont {Li}}, \bibinfo {author}
  {\bibfnamefont {M.~B.}\ \bibnamefont {Stone}}, \bibinfo {author}
  {\bibfnamefont {G.~E.}\ \bibnamefont {Granroth}}, \bibinfo {author}
  {\bibfnamefont {M.~D.}\ \bibnamefont {Lumsden}}, \bibinfo {author}
  {\bibfnamefont {Y.}~\bibnamefont {Yiu}}, \bibinfo {author} {\bibfnamefont
  {J.}~\bibnamefont {Knolle}}, \bibinfo {author} {\bibfnamefont
  {S.}~\bibnamefont {Bhattacharjee}}, \bibinfo {author} {\bibfnamefont {D.~L.}\
  \bibnamefont {Kovrizhin}}, \bibinfo {author} {\bibfnamefont {R.}~\bibnamefont
  {Moessner}}, \bibinfo {author} {\bibfnamefont {D.~A.}\ \bibnamefont
  {Tennant}}, \bibinfo {author} {\bibfnamefont {D.~G.}\ \bibnamefont
  {Mandrus}}, \ and\ \bibinfo {author} {\bibfnamefont {S.~E.}\ \bibnamefont
  {Nagler}},\ }\href {\doibase 10.1038/nmat4604} {\bibfield  {journal}
  {\bibinfo  {journal} {Nat. Mater.}\ }\textbf {\bibinfo {volume} {15}},\
  \bibinfo {pages} {733} (\bibinfo {year} {2016})}\BibitemShut {NoStop}%
\bibitem [{\citenamefont {Williams}\ \emph {et~al.}(2016)\citenamefont
  {Williams}, \citenamefont {Johnson}, \citenamefont {Freund}, \citenamefont
  {Choi}, \citenamefont {Jesche}, \citenamefont {Kimchi}, \citenamefont
  {Manni}, \citenamefont {Bombardi}, \citenamefont {Manuel}, \citenamefont
  {Gegenwart},\ and\ \citenamefont {Coldea}}]{Steph:Li213}%
  \BibitemOpen
  \bibfield  {author} {\bibinfo {author} {\bibfnamefont {S.~C.}\ \bibnamefont
  {Williams}}, \bibinfo {author} {\bibfnamefont {R.~D.}\ \bibnamefont
  {Johnson}}, \bibinfo {author} {\bibfnamefont {F.}~\bibnamefont {Freund}},
  \bibinfo {author} {\bibfnamefont {S.}~\bibnamefont {Choi}}, \bibinfo {author}
  {\bibfnamefont {A.}~\bibnamefont {Jesche}}, \bibinfo {author} {\bibfnamefont
  {I.}~\bibnamefont {Kimchi}}, \bibinfo {author} {\bibfnamefont
  {S.}~\bibnamefont {Manni}}, \bibinfo {author} {\bibfnamefont
  {A.}~\bibnamefont {Bombardi}}, \bibinfo {author} {\bibfnamefont
  {P.}~\bibnamefont {Manuel}}, \bibinfo {author} {\bibfnamefont
  {P.}~\bibnamefont {Gegenwart}}, \ and\ \bibinfo {author} {\bibfnamefont
  {R.}~\bibnamefont {Coldea}},\ }\href {\doibase 10.1103/PhysRevB.93.195158}
  {\bibfield  {journal} {\bibinfo  {journal} {Phys. Rev. B}\ }\textbf {\bibinfo
  {volume} {93}},\ \bibinfo {pages} {195158} (\bibinfo {year}
  {2016})}\BibitemShut {NoStop}%
\bibitem [{\citenamefont {Biffin}\ \emph
  {et~al.}(2014{\natexlab{a}})\citenamefont {Biffin}, \citenamefont {Johnson},
  \citenamefont {Choi}, \citenamefont {Freund}, \citenamefont {Manni},
  \citenamefont {Bombardi}, \citenamefont {Manuel}, \citenamefont {Gegenwart},\
  and\ \citenamefont {Coldea}}]{alun:beta:Li213}%
  \BibitemOpen
  \bibfield  {author} {\bibinfo {author} {\bibfnamefont {A.}~\bibnamefont
  {Biffin}}, \bibinfo {author} {\bibfnamefont {R.~D.}\ \bibnamefont {Johnson}},
  \bibinfo {author} {\bibfnamefont {S.}~\bibnamefont {Choi}}, \bibinfo {author}
  {\bibfnamefont {F.}~\bibnamefont {Freund}}, \bibinfo {author} {\bibfnamefont
  {S.}~\bibnamefont {Manni}}, \bibinfo {author} {\bibfnamefont
  {A.}~\bibnamefont {Bombardi}}, \bibinfo {author} {\bibfnamefont
  {P.}~\bibnamefont {Manuel}}, \bibinfo {author} {\bibfnamefont
  {P.}~\bibnamefont {Gegenwart}}, \ and\ \bibinfo {author} {\bibfnamefont
  {R.}~\bibnamefont {Coldea}},\ }\href@noop {} {\bibfield  {journal} {\bibinfo
  {journal} {Phys. Rev. B}\ }\textbf {\bibinfo {volume} {90}},\ \bibinfo
  {pages} {205116} (\bibinfo {year} {2014}{\natexlab{a}})}\BibitemShut
  {NoStop}%
\bibitem [{\citenamefont {Biffin}\ \emph
  {et~al.}(2014{\natexlab{b}})\citenamefont {Biffin}, \citenamefont {Johnson},
  \citenamefont {Kimchi}, \citenamefont {Morris}, \citenamefont {Bombardi},
  \citenamefont {Analytis}, \citenamefont {Vishwanath},\ and\ \citenamefont
  {Coldea}}]{alun:gamma:Li213}%
  \BibitemOpen
  \bibfield  {author} {\bibinfo {author} {\bibfnamefont {A.}~\bibnamefont
  {Biffin}}, \bibinfo {author} {\bibfnamefont {R.~D.}\ \bibnamefont {Johnson}},
  \bibinfo {author} {\bibfnamefont {I.}~\bibnamefont {Kimchi}}, \bibinfo
  {author} {\bibfnamefont {R.}~\bibnamefont {Morris}}, \bibinfo {author}
  {\bibfnamefont {A.}~\bibnamefont {Bombardi}}, \bibinfo {author}
  {\bibfnamefont {J.~G.}\ \bibnamefont {Analytis}}, \bibinfo {author}
  {\bibfnamefont {A.}~\bibnamefont {Vishwanath}}, \ and\ \bibinfo {author}
  {\bibfnamefont {R.}~\bibnamefont {Coldea}},\ }\href@noop {} {\bibfield
  {journal} {\bibinfo  {journal} {Phys. Rev. Lett.}\ }\textbf {\bibinfo
  {volume} {113}},\ \bibinfo {pages} {197201} (\bibinfo {year}
  {2014}{\natexlab{b}})}\BibitemShut {NoStop}%
\bibitem [{\citenamefont {Kitagawa}\ \emph {et~al.}(2018)\citenamefont
  {Kitagawa}, \citenamefont {Takayama}, \citenamefont {Matsumoto},
  \citenamefont {Kato}, \citenamefont {Takano}, \citenamefont {Kishimoto},
  \citenamefont {Bette}, \citenamefont {Dinnebier}, \citenamefont {Jackeli},\
  and\ \citenamefont {Takagi}}]{Takagi2017}%
  \BibitemOpen
  \bibfield  {author} {\bibinfo {author} {\bibfnamefont {K.}~\bibnamefont
  {Kitagawa}}, \bibinfo {author} {\bibfnamefont {T.}~\bibnamefont {Takayama}},
  \bibinfo {author} {\bibfnamefont {Y.}~\bibnamefont {Matsumoto}}, \bibinfo
  {author} {\bibfnamefont {A.}~\bibnamefont {Kato}}, \bibinfo {author}
  {\bibfnamefont {R.}~\bibnamefont {Takano}}, \bibinfo {author} {\bibfnamefont
  {Y.}~\bibnamefont {Kishimoto}}, \bibinfo {author} {\bibfnamefont
  {S.}~\bibnamefont {Bette}}, \bibinfo {author} {\bibfnamefont
  {R.}~\bibnamefont {Dinnebier}}, \bibinfo {author} {\bibfnamefont
  {G.}~\bibnamefont {Jackeli}}, \ and\ \bibinfo {author} {\bibfnamefont
  {H.}~\bibnamefont {Takagi}},\ }\href {\doibase 10.1038/nature25482}
  {\bibfield  {journal} {\bibinfo  {journal} {Nature}\ }\textbf {\bibinfo
  {volume} {554}},\ \bibinfo {pages} {341} (\bibinfo {year}
  {2018})}\BibitemShut {NoStop}%
\bibitem [{\citenamefont {Veiga}\ \emph {et~al.}(2017)\citenamefont {Veiga},
  \citenamefont {Etter}, \citenamefont {Glazyrin}, \citenamefont {Sun},
  \citenamefont {Escanhoela}, \citenamefont {Fabbris}, \citenamefont
  {Mardegan}, \citenamefont {Malavi}, \citenamefont {Deng}, \citenamefont
  {Stavropoulos}, \citenamefont {Kee}, \citenamefont {Yang}, \citenamefont {van
  Veenendaal}, \citenamefont {Schilling}, \citenamefont {Takayama},
  \citenamefont {Takagi},\ and\ \citenamefont {Haskel}}]{Veiga2017}%
  \BibitemOpen
  \bibfield  {author} {\bibinfo {author} {\bibfnamefont {L.~S.~I.}\
  \bibnamefont {Veiga}}, \bibinfo {author} {\bibfnamefont {M.}~\bibnamefont
  {Etter}}, \bibinfo {author} {\bibfnamefont {K.}~\bibnamefont {Glazyrin}},
  \bibinfo {author} {\bibfnamefont {F.}~\bibnamefont {Sun}}, \bibinfo {author}
  {\bibfnamefont {C.~A.}\ \bibnamefont {Escanhoela}}, \bibinfo {author}
  {\bibfnamefont {G.}~\bibnamefont {Fabbris}}, \bibinfo {author} {\bibfnamefont
  {J.~R.~L.}\ \bibnamefont {Mardegan}}, \bibinfo {author} {\bibfnamefont
  {P.~S.}\ \bibnamefont {Malavi}}, \bibinfo {author} {\bibfnamefont
  {Y.}~\bibnamefont {Deng}}, \bibinfo {author} {\bibfnamefont {P.~P.}\
  \bibnamefont {Stavropoulos}}, \bibinfo {author} {\bibfnamefont {H.-Y.}\
  \bibnamefont {Kee}}, \bibinfo {author} {\bibfnamefont {W.~G.}\ \bibnamefont
  {Yang}}, \bibinfo {author} {\bibfnamefont {M.}~\bibnamefont {van
  Veenendaal}}, \bibinfo {author} {\bibfnamefont {J.~S.}\ \bibnamefont
  {Schilling}}, \bibinfo {author} {\bibfnamefont {T.}~\bibnamefont {Takayama}},
  \bibinfo {author} {\bibfnamefont {H.}~\bibnamefont {Takagi}}, \ and\ \bibinfo
  {author} {\bibfnamefont {D.}~\bibnamefont {Haskel}},\ }\href {\doibase
  10.1103/PhysRevB.96.140402} {\bibfield  {journal} {\bibinfo  {journal} {Phys.
  Rev. B}\ }\textbf {\bibinfo {volume} {96}},\ \bibinfo {pages} {140402}
  (\bibinfo {year} {2017})}\BibitemShut {NoStop}%
\bibitem [{\citenamefont {Majumder}\ \emph {et~al.}(2018)\citenamefont
  {Majumder}, \citenamefont {Manna}, \citenamefont {Simutis}, \citenamefont
  {Orain}, \citenamefont {Dey}, \citenamefont {Freund}, \citenamefont {Jesche},
  \citenamefont {Khasanov}, \citenamefont {Biswas}, \citenamefont {Bykova},
  \citenamefont {Dubrovinskaia}, \citenamefont {Dubrovinsky}, \citenamefont
  {Yadav}, \citenamefont {Hozoi}, \citenamefont {Nishimoto}, \citenamefont
  {Tsirlin},\ and\ \citenamefont {Gegenwart}}]{Majumder2018}%
  \BibitemOpen
  \bibfield  {author} {\bibinfo {author} {\bibfnamefont {M.}~\bibnamefont
  {Majumder}}, \bibinfo {author} {\bibfnamefont {R.~S.}\ \bibnamefont {Manna}},
  \bibinfo {author} {\bibfnamefont {G.}~\bibnamefont {Simutis}}, \bibinfo
  {author} {\bibfnamefont {J.~C.}\ \bibnamefont {Orain}}, \bibinfo {author}
  {\bibfnamefont {T.}~\bibnamefont {Dey}}, \bibinfo {author} {\bibfnamefont
  {F.}~\bibnamefont {Freund}}, \bibinfo {author} {\bibfnamefont
  {A.}~\bibnamefont {Jesche}}, \bibinfo {author} {\bibfnamefont
  {R.}~\bibnamefont {Khasanov}}, \bibinfo {author} {\bibfnamefont {P.~K.}\
  \bibnamefont {Biswas}}, \bibinfo {author} {\bibfnamefont {E.}~\bibnamefont
  {Bykova}}, \bibinfo {author} {\bibfnamefont {N.}~\bibnamefont
  {Dubrovinskaia}}, \bibinfo {author} {\bibfnamefont {L.~S.}\ \bibnamefont
  {Dubrovinsky}}, \bibinfo {author} {\bibfnamefont {R.}~\bibnamefont {Yadav}},
  \bibinfo {author} {\bibfnamefont {L.}~\bibnamefont {Hozoi}}, \bibinfo
  {author} {\bibfnamefont {S.}~\bibnamefont {Nishimoto}}, \bibinfo {author}
  {\bibfnamefont {A.~A.}\ \bibnamefont {Tsirlin}}, \ and\ \bibinfo {author}
  {\bibfnamefont {P.}~\bibnamefont {Gegenwart}},\ }\href@noop {} {\bibfield
  {journal} {\bibinfo  {journal} {Phys. Rev. Lett.}\ }\textbf {\bibinfo
  {volume} {120}},\ \bibinfo {pages} {237202} (\bibinfo {year}
  {2018})}\BibitemShut {NoStop}%
\bibitem [{\citenamefont {Takayama}\ \emph {et~al.}()\citenamefont {Takayama},
  \citenamefont {Krajewska}, \citenamefont {Gibbs}, \citenamefont {Yaresko},
  \citenamefont {Ishii}, \citenamefont {Yamaoka}, \citenamefont {Ishii},
  \citenamefont {Hiraoka}, \citenamefont {Funnell}, \citenamefont {Bull},\ and\
  \citenamefont {Takagi}}]{Takayama2017}%
  \BibitemOpen
  \bibfield  {author} {\bibinfo {author} {\bibfnamefont {T.}~\bibnamefont
  {Takayama}}, \bibinfo {author} {\bibfnamefont {A.}~\bibnamefont {Krajewska}},
  \bibinfo {author} {\bibfnamefont {A.~S.}\ \bibnamefont {Gibbs}}, \bibinfo
  {author} {\bibfnamefont {A.~N.}\ \bibnamefont {Yaresko}}, \bibinfo {author}
  {\bibfnamefont {H.}~\bibnamefont {Ishii}}, \bibinfo {author} {\bibfnamefont
  {H.}~\bibnamefont {Yamaoka}}, \bibinfo {author} {\bibfnamefont
  {K.}~\bibnamefont {Ishii}}, \bibinfo {author} {\bibfnamefont
  {N.}~\bibnamefont {Hiraoka}}, \bibinfo {author} {\bibfnamefont {N.~P.}\
  \bibnamefont {Funnell}}, \bibinfo {author} {\bibfnamefont {C.~L.}\
  \bibnamefont {Bull}}, \ and\ \bibinfo {author} {\bibfnamefont
  {H.}~\bibnamefont {Takagi}},\ }\href@noop {} {\bibinfo  {journal}
  {arXiv:1808.05494}\ }\BibitemShut {NoStop}%
\bibitem [{\citenamefont {Choi}\ and\ \citenamefont
  {et~al.}()}]{bLIO_HP_Raman2018}%
  \BibitemOpen
\bibfield  {journal} {  }\bibfield  {author} {\bibinfo {author} {\bibfnamefont
  {S.}~\bibnamefont {Choi}}\ and\ \bibinfo {author} {\bibnamefont {et~al.}},\
  }\href@noop {} {\bibinfo  {journal} {(unpublished)}\ }\BibitemShut {NoStop}%
\bibitem [{\citenamefont {Breznay}\ \emph {et~al.}(2017)\citenamefont
  {Breznay}, \citenamefont {Ruiz}, \citenamefont {Frano}, \citenamefont {Bi},
  \citenamefont {Birgeneau}, \citenamefont {Haskel},\ and\ \citenamefont
  {Analytis}}]{Breznay2017}%
  \BibitemOpen
\bibfield  {journal} {  }\bibfield  {author} {\bibinfo {author} {\bibfnamefont
  {N.~P.}\ \bibnamefont {Breznay}}, \bibinfo {author} {\bibfnamefont
  {A.}~\bibnamefont {Ruiz}}, \bibinfo {author} {\bibfnamefont {A.}~\bibnamefont
  {Frano}}, \bibinfo {author} {\bibfnamefont {W.}~\bibnamefont {Bi}}, \bibinfo
  {author} {\bibfnamefont {R.~J.}\ \bibnamefont {Birgeneau}}, \bibinfo {author}
  {\bibfnamefont {D.}~\bibnamefont {Haskel}}, \ and\ \bibinfo {author}
  {\bibfnamefont {J.~G.}\ \bibnamefont {Analytis}},\ }\href {\doibase
  10.1103/PhysRevB.96.020402} {\bibfield  {journal} {\bibinfo  {journal} {Phys.
  Rev. B}\ }\textbf {\bibinfo {volume} {96}},\ \bibinfo {pages} {020402}
  (\bibinfo {year} {2017})}\BibitemShut {NoStop}%
\bibitem [{\citenamefont {Banerjee}\ \emph {et~al.}(2017)\citenamefont
  {Banerjee}, \citenamefont {Yan}, \citenamefont {Knolle}, \citenamefont
  {Bridges}, \citenamefont {Stone}, \citenamefont {Lumsden}, \citenamefont
  {Mandrus}, \citenamefont {Tennant}, \citenamefont {Moessner},\ and\
  \citenamefont {Nagler}}]{Banerjee2017}%
  \BibitemOpen
  \bibfield  {author} {\bibinfo {author} {\bibfnamefont {A.}~\bibnamefont
  {Banerjee}}, \bibinfo {author} {\bibfnamefont {J.}~\bibnamefont {Yan}},
  \bibinfo {author} {\bibfnamefont {J.}~\bibnamefont {Knolle}}, \bibinfo
  {author} {\bibfnamefont {C.~A.}\ \bibnamefont {Bridges}}, \bibinfo {author}
  {\bibfnamefont {M.~B.}\ \bibnamefont {Stone}}, \bibinfo {author}
  {\bibfnamefont {M.~D.}\ \bibnamefont {Lumsden}}, \bibinfo {author}
  {\bibfnamefont {D.~G.}\ \bibnamefont {Mandrus}}, \bibinfo {author}
  {\bibfnamefont {D.~A.}\ \bibnamefont {Tennant}}, \bibinfo {author}
  {\bibfnamefont {R.}~\bibnamefont {Moessner}}, \ and\ \bibinfo {author}
  {\bibfnamefont {S.~E.}\ \bibnamefont {Nagler}},\ }\href {\doibase
  10.1126/science.aah6015} {\bibfield  {journal} {\bibinfo  {journal}
  {Science}\ }\textbf {\bibinfo {volume} {356}},\ \bibinfo {pages} {1055}
  (\bibinfo {year} {2017})}\BibitemShut {NoStop}%
\bibitem [{\citenamefont {Do}\ \emph {et~al.}(2017)\citenamefont {Do},
  \citenamefont {Park}, \citenamefont {Yoshitake}, \citenamefont {Nasu},
  \citenamefont {Motome}, \citenamefont {Kwon}, \citenamefont {Adroja},
  \citenamefont {Voneshen}, \citenamefont {Kim}, \citenamefont {Jang},
  \citenamefont {Park}, \citenamefont {Choi},\ and\ \citenamefont
  {Ji}}]{Do2017}%
  \BibitemOpen
  \bibfield  {author} {\bibinfo {author} {\bibfnamefont {S.-H.}\ \bibnamefont
  {Do}}, \bibinfo {author} {\bibfnamefont {S.-Y.}\ \bibnamefont {Park}},
  \bibinfo {author} {\bibfnamefont {J.}~\bibnamefont {Yoshitake}}, \bibinfo
  {author} {\bibfnamefont {J.}~\bibnamefont {Nasu}}, \bibinfo {author}
  {\bibfnamefont {Y.}~\bibnamefont {Motome}}, \bibinfo {author} {\bibfnamefont
  {Y.~S.}\ \bibnamefont {Kwon}}, \bibinfo {author} {\bibfnamefont {D.~T.}\
  \bibnamefont {Adroja}}, \bibinfo {author} {\bibfnamefont {D.~J.}\
  \bibnamefont {Voneshen}}, \bibinfo {author} {\bibfnamefont {K.}~\bibnamefont
  {Kim}}, \bibinfo {author} {\bibfnamefont {T.-H.}\ \bibnamefont {Jang}},
  \bibinfo {author} {\bibfnamefont {J.-H.}\ \bibnamefont {Park}}, \bibinfo
  {author} {\bibfnamefont {K.-Y.}\ \bibnamefont {Choi}}, \ and\ \bibinfo
  {author} {\bibfnamefont {S.}~\bibnamefont {Ji}},\ }\href {\doibase
  10.1038/nphys4264} {\bibfield  {journal} {\bibinfo  {journal} {Nature Phys.}\
  }\textbf {\bibinfo {volume} {13}},\ \bibinfo {pages} {1079} (\bibinfo {year}
  {2017})}\BibitemShut {NoStop}%
\bibitem [{\citenamefont {Gretarsson}\ \emph {et~al.}(2013)\citenamefont
  {Gretarsson}, \citenamefont {Clancy}, \citenamefont {Singh}, \citenamefont
  {Gegenwart}, \citenamefont {Hill}, \citenamefont {Kim}, \citenamefont
  {Upton}, \citenamefont {Said}, \citenamefont {Casa}, \citenamefont {Gog},\
  and\ \citenamefont {Kim}}]{Na213_RIXS_GH_2013}%
  \BibitemOpen
  \bibfield  {author} {\bibinfo {author} {\bibfnamefont {H.}~\bibnamefont
  {Gretarsson}}, \bibinfo {author} {\bibfnamefont {J.~P.}\ \bibnamefont
  {Clancy}}, \bibinfo {author} {\bibfnamefont {Y.}~\bibnamefont {Singh}},
  \bibinfo {author} {\bibfnamefont {P.}~\bibnamefont {Gegenwart}}, \bibinfo
  {author} {\bibfnamefont {J.~P.}\ \bibnamefont {Hill}}, \bibinfo {author}
  {\bibfnamefont {J.}~\bibnamefont {Kim}}, \bibinfo {author} {\bibfnamefont
  {M.~H.}\ \bibnamefont {Upton}}, \bibinfo {author} {\bibfnamefont {A.~H.}\
  \bibnamefont {Said}}, \bibinfo {author} {\bibfnamefont {D.}~\bibnamefont
  {Casa}}, \bibinfo {author} {\bibfnamefont {T.}~\bibnamefont {Gog}}, \ and\
  \bibinfo {author} {\bibfnamefont {Y.-J.}\ \bibnamefont {Kim}},\ }\href@noop
  {} {\bibfield  {journal} {\bibinfo  {journal} {Phys. Rev. B}\ }\textbf
  {\bibinfo {volume} {87}},\ \bibinfo {pages} {220407} (\bibinfo {year}
  {2013})}\BibitemShut {NoStop}%
\bibitem [{\citenamefont {Kimchi}\ \emph {et~al.}(2015)\citenamefont {Kimchi},
  \citenamefont {Coldea},\ and\ \citenamefont {Vishwanath}}]{Kimchi:abcLi213}%
  \BibitemOpen
  \bibfield  {author} {\bibinfo {author} {\bibfnamefont {I.}~\bibnamefont
  {Kimchi}}, \bibinfo {author} {\bibfnamefont {R.}~\bibnamefont {Coldea}}, \
  and\ \bibinfo {author} {\bibfnamefont {A.}~\bibnamefont {Vishwanath}},\
  }\href {\doibase 10.1103/PhysRevB.91.245134} {\bibfield  {journal} {\bibinfo
  {journal} {Phys. Rev. B}\ }\textbf {\bibinfo {volume} {91}},\ \bibinfo
  {pages} {245134} (\bibinfo {year} {2015})}\BibitemShut {NoStop}%
\bibitem [{\citenamefont {Lee}\ and\ \citenamefont {Kim}(2015)}]{Lee2015}%
  \BibitemOpen
  \bibfield  {author} {\bibinfo {author} {\bibfnamefont {E.~K.-H.}\
  \bibnamefont {Lee}}\ and\ \bibinfo {author} {\bibfnamefont {Y.~B.}\
  \bibnamefont {Kim}},\ }\href {\doibase 10.1103/PhysRevB.91.064407} {\bibfield
   {journal} {\bibinfo  {journal} {Phys. Rev. B}\ }\textbf {\bibinfo {volume}
  {91}},\ \bibinfo {pages} {064407} (\bibinfo {year} {2015})}\BibitemShut
  {NoStop}%
\bibitem [{\citenamefont {Lee}\ \emph {et~al.}(2016)\citenamefont {Lee},
  \citenamefont {Rau},\ and\ \citenamefont {Kim}}]{Lee2016}%
  \BibitemOpen
  \bibfield  {author} {\bibinfo {author} {\bibfnamefont {E.~K.-H.}\
  \bibnamefont {Lee}}, \bibinfo {author} {\bibfnamefont {J.~G.}\ \bibnamefont
  {Rau}}, \ and\ \bibinfo {author} {\bibfnamefont {Y.~B.}\ \bibnamefont
  {Kim}},\ }\href {\doibase 10.1103/PhysRevB.93.184420} {\bibfield  {journal}
  {\bibinfo  {journal} {Phys. Rev. B}\ }\textbf {\bibinfo {volume} {93}},\
  \bibinfo {pages} {184420} (\bibinfo {year} {2016})}\BibitemShut {NoStop}%
\bibitem [{\citenamefont {Kimchi}\ and\ \citenamefont
  {Coldea}(2016)}]{Kimchi2016}%
  \BibitemOpen
  \bibfield  {author} {\bibinfo {author} {\bibfnamefont {I.}~\bibnamefont
  {Kimchi}}\ and\ \bibinfo {author} {\bibfnamefont {R.}~\bibnamefont
  {Coldea}},\ }\href {\doibase 10.1103/PhysRevB.94.201110} {\bibfield
  {journal} {\bibinfo  {journal} {Phys. Rev. B}\ }\textbf {\bibinfo {volume}
  {94}},\ \bibinfo {pages} {201110} (\bibinfo {year} {2016})}\BibitemShut
  {NoStop}%
\bibitem [{\citenamefont {Ducatman}\ \emph {et~al.}(2018)\citenamefont
  {Ducatman}, \citenamefont {Rousochatzakis},\ and\ \citenamefont
  {Perkins}}]{Perkins2017}%
  \BibitemOpen
  \bibfield  {author} {\bibinfo {author} {\bibfnamefont {S.}~\bibnamefont
  {Ducatman}}, \bibinfo {author} {\bibfnamefont {I.}~\bibnamefont
  {Rousochatzakis}}, \ and\ \bibinfo {author} {\bibfnamefont {N.~B.}\
  \bibnamefont {Perkins}},\ }\href {\doibase 10.1103/PhysRevB.97.125125}
  {\bibfield  {journal} {\bibinfo  {journal} {Phys. Rev. B}\ }\textbf {\bibinfo
  {volume} {97}},\ \bibinfo {pages} {125125} (\bibinfo {year}
  {2018})}\BibitemShut {NoStop}%
\bibitem [{\citenamefont {Sears}\ \emph {et~al.}(2017)\citenamefont {Sears},
  \citenamefont {Zhao}, \citenamefont {Xu}, \citenamefont {Lynn},\ and\
  \citenamefont {Kim}}]{Sears2017}%
  \BibitemOpen
  \bibfield  {author} {\bibinfo {author} {\bibfnamefont {J.~A.}\ \bibnamefont
  {Sears}}, \bibinfo {author} {\bibfnamefont {Y.}~\bibnamefont {Zhao}},
  \bibinfo {author} {\bibfnamefont {Z.}~\bibnamefont {Xu}}, \bibinfo {author}
  {\bibfnamefont {J.~W.}\ \bibnamefont {Lynn}}, \ and\ \bibinfo {author}
  {\bibfnamefont {Y.-J.}\ \bibnamefont {Kim}},\ }\href {\doibase
  10.1103/PhysRevB.95.180411} {\bibfield  {journal} {\bibinfo  {journal} {Phys.
  Rev. B}\ }\textbf {\bibinfo {volume} {95}},\ \bibinfo {pages} {180411}
  (\bibinfo {year} {2017})}\BibitemShut {NoStop}%
\bibitem [{\citenamefont {Ruiz}\ \emph {et~al.}(2017)\citenamefont {Ruiz},
  \citenamefont {Frano}, \citenamefont {Breznay}, \citenamefont {Kimchi},
  \citenamefont {Helm}, \citenamefont {Oswald}, \citenamefont {Chan},
  \citenamefont {Birgeneau}, \citenamefont {Islam},\ and\ \citenamefont
  {Analytis}}]{Ruiz2017}%
  \BibitemOpen
  \bibfield  {author} {\bibinfo {author} {\bibfnamefont {A.}~\bibnamefont
  {Ruiz}}, \bibinfo {author} {\bibfnamefont {A.}~\bibnamefont {Frano}},
  \bibinfo {author} {\bibfnamefont {N.~P.}\ \bibnamefont {Breznay}}, \bibinfo
  {author} {\bibfnamefont {I.}~\bibnamefont {Kimchi}}, \bibinfo {author}
  {\bibfnamefont {T.}~\bibnamefont {Helm}}, \bibinfo {author} {\bibfnamefont
  {I.}~\bibnamefont {Oswald}}, \bibinfo {author} {\bibfnamefont {J.~Y.}\
  \bibnamefont {Chan}}, \bibinfo {author} {\bibfnamefont {R.~J.}\ \bibnamefont
  {Birgeneau}}, \bibinfo {author} {\bibfnamefont {Z.}~\bibnamefont {Islam}}, \
  and\ \bibinfo {author} {\bibfnamefont {J.~G.}\ \bibnamefont {Analytis}},\
  }\href {\doibase 10.1038/s41467-017-01071-9} {\bibfield  {journal} {\bibinfo
  {journal} {Nat. Commun.}\ }\textbf {\bibinfo {volume} {8}},\ \bibinfo {pages}
  {961} (\bibinfo {year} {2017})}\BibitemShut {NoStop}%
\bibitem [{\citenamefont {Modic}\ \emph {et~al.}(2017)\citenamefont {Modic},
  \citenamefont {Ramshaw}, \citenamefont {Betts}, \citenamefont {Breznay},
  \citenamefont {Analytis}, \citenamefont {McDonald},\ and\ \citenamefont
  {Shekhter}}]{Modic2017}%
  \BibitemOpen
  \bibfield  {author} {\bibinfo {author} {\bibfnamefont {K.~A.}\ \bibnamefont
  {Modic}}, \bibinfo {author} {\bibfnamefont {B.~J.}\ \bibnamefont {Ramshaw}},
  \bibinfo {author} {\bibfnamefont {J.~B.}\ \bibnamefont {Betts}}, \bibinfo
  {author} {\bibfnamefont {N.~P.}\ \bibnamefont {Breznay}}, \bibinfo {author}
  {\bibfnamefont {J.~G.}\ \bibnamefont {Analytis}}, \bibinfo {author}
  {\bibfnamefont {R.~D.}\ \bibnamefont {McDonald}}, \ and\ \bibinfo {author}
  {\bibfnamefont {A.}~\bibnamefont {Shekhter}},\ }\href {\doibase
  10.1038/s41467-017-00264-6} {\bibfield  {journal} {\bibinfo  {journal} {Nat.
  Commun.}\ }\textbf {\bibinfo {volume} {8}},\ \bibinfo {pages} {180} (\bibinfo
  {year} {2017})}\BibitemShut {NoStop}%
\bibitem [{\citenamefont {Banerjee}\ \emph {et~al.}(2018)\citenamefont
  {Banerjee}, \citenamefont {Lampen-Kelley}, \citenamefont {Knolle},
  \citenamefont {Balz}, \citenamefont {Aczel}, \citenamefont {Winn},
  \citenamefont {Liu}, \citenamefont {Pajerowski}, \citenamefont {Yan},
  \citenamefont {Bridges}, \citenamefont {Savici}, \citenamefont {Chakoumakos},
  \citenamefont {Lumsden}, \citenamefont {Tennant}, \citenamefont {Moessner},
  \citenamefont {Mandrus},\ and\ \citenamefont {Nagler}}]{BanerjeeB}%
  \BibitemOpen
  \bibfield  {author} {\bibinfo {author} {\bibfnamefont {A.}~\bibnamefont
  {Banerjee}}, \bibinfo {author} {\bibfnamefont {P.}~\bibnamefont
  {Lampen-Kelley}}, \bibinfo {author} {\bibfnamefont {J.}~\bibnamefont
  {Knolle}}, \bibinfo {author} {\bibfnamefont {C.}~\bibnamefont {Balz}},
  \bibinfo {author} {\bibfnamefont {A.}~\bibnamefont {Aczel}}, \bibinfo
  {author} {\bibfnamefont {B.}~\bibnamefont {Winn}}, \bibinfo {author}
  {\bibfnamefont {Y.}~\bibnamefont {Liu}}, \bibinfo {author} {\bibfnamefont
  {D.}~\bibnamefont {Pajerowski}}, \bibinfo {author} {\bibfnamefont {J.-Q.}\
  \bibnamefont {Yan}}, \bibinfo {author} {\bibfnamefont {C.}~\bibnamefont
  {Bridges}}, \bibinfo {author} {\bibfnamefont {A.}~\bibnamefont {Savici}},
  \bibinfo {author} {\bibfnamefont {B.}~\bibnamefont {Chakoumakos}}, \bibinfo
  {author} {\bibfnamefont {M.}~\bibnamefont {Lumsden}}, \bibinfo {author}
  {\bibfnamefont {D.}~\bibnamefont {Tennant}}, \bibinfo {author} {\bibfnamefont
  {R.}~\bibnamefont {Moessner}}, \bibinfo {author} {\bibfnamefont
  {D.}~\bibnamefont {Mandrus}}, \ and\ \bibinfo {author} {\bibfnamefont
  {S.}~\bibnamefont {Nagler}},\ }\href {\doibase 10.1038/s41535-018-0079-2}
  {\bibfield  {journal} {\bibinfo  {journal} {npj Quantum Materials}\ }\textbf
  {\bibinfo {volume} {3}},\ \bibinfo {pages} {Article number: 8} (\bibinfo
  {year} {2018})}\BibitemShut {NoStop}%
\bibitem [{\citenamefont {Winter}\ \emph {et~al.}(2018)\citenamefont {Winter},
  \citenamefont {Riedl}, \citenamefont {Kaib}, \citenamefont {Coldea},\ and\
  \citenamefont {Valent\'{\i}}}]{WinterB}%
  \BibitemOpen
  \bibfield  {author} {\bibinfo {author} {\bibfnamefont {S.~M.}\ \bibnamefont
  {Winter}}, \bibinfo {author} {\bibfnamefont {K.}~\bibnamefont {Riedl}},
  \bibinfo {author} {\bibfnamefont {D.}~\bibnamefont {Kaib}}, \bibinfo {author}
  {\bibfnamefont {R.}~\bibnamefont {Coldea}}, \ and\ \bibinfo {author}
  {\bibfnamefont {R.}~\bibnamefont {Valent\'{\i}}},\ }\href {\doibase
  10.1103/PhysRevLett.120.077203} {\bibfield  {journal} {\bibinfo  {journal}
  {Phys. Rev. Lett.}\ }\textbf {\bibinfo {volume} {120}},\ \bibinfo {pages}
  {077203} (\bibinfo {year} {2018})}\BibitemShut {NoStop}%
\bibitem [{\citenamefont {Lampen-Kelley}\ \emph {et~al.}()\citenamefont
  {Lampen-Kelley}, \citenamefont {Janssen}, \citenamefont {Andrade},
  \citenamefont {Rachel}, \citenamefont {Yan}, \citenamefont {Balz},
  \citenamefont {Mandrus}, \citenamefont {Nagler},\ and\ \citenamefont
  {Vojta}}]{Lampen2018}%
  \BibitemOpen
  \bibfield  {author} {\bibinfo {author} {\bibfnamefont {P.}~\bibnamefont
  {Lampen-Kelley}}, \bibinfo {author} {\bibfnamefont {L.}~\bibnamefont
  {Janssen}}, \bibinfo {author} {\bibfnamefont {E.~C.}\ \bibnamefont
  {Andrade}}, \bibinfo {author} {\bibfnamefont {S.}~\bibnamefont {Rachel}},
  \bibinfo {author} {\bibfnamefont {J.-Q.}\ \bibnamefont {Yan}}, \bibinfo
  {author} {\bibfnamefont {C.}~\bibnamefont {Balz}}, \bibinfo {author}
  {\bibfnamefont {D.~G.}\ \bibnamefont {Mandrus}}, \bibinfo {author}
  {\bibfnamefont {S.~E.}\ \bibnamefont {Nagler}}, \ and\ \bibinfo {author}
  {\bibfnamefont {M.}~\bibnamefont {Vojta}},\ }\href@noop {} {\bibinfo
  {journal} {arXiv:1807.06192}\ }\BibitemShut {NoStop}%
\bibitem [{\citenamefont {Blundell}(1999)}]{SB_muSR}%
  \BibitemOpen
\bibfield  {journal} {  }\bibfield  {author} {\bibinfo {author} {\bibfnamefont
  {S.~J.}\ \bibnamefont {Blundell}},\ }\href@noop {} {\bibfield  {journal}
  {\bibinfo  {journal} {Contemp. Phys.}\ }\textbf {\bibinfo {volume} {40}},\
  \bibinfo {pages} {175} (\bibinfo {year} {1999})}\BibitemShut {NoStop}%
\bibitem [{\citenamefont {Sugiyama}\ \emph {et~al.}(2004)\citenamefont
  {Sugiyama}, \citenamefont {Brewer}, \citenamefont {Ansaldo}, \citenamefont
  {Hitti}, \citenamefont {Mikami}, \citenamefont {Mori},\ and\ \citenamefont
  {Sasaki}}]{Sugiyama:muSR:2004}%
  \BibitemOpen
  \bibfield  {author} {\bibinfo {author} {\bibfnamefont {J.}~\bibnamefont
  {Sugiyama}}, \bibinfo {author} {\bibfnamefont {J.~H.}\ \bibnamefont
  {Brewer}}, \bibinfo {author} {\bibfnamefont {E.~J.}\ \bibnamefont {Ansaldo}},
  \bibinfo {author} {\bibfnamefont {B.}~\bibnamefont {Hitti}}, \bibinfo
  {author} {\bibfnamefont {M.}~\bibnamefont {Mikami}}, \bibinfo {author}
  {\bibfnamefont {Y.}~\bibnamefont {Mori}}, \ and\ \bibinfo {author}
  {\bibfnamefont {T.}~\bibnamefont {Sasaki}},\ }\href@noop {} {\bibfield
  {journal} {\bibinfo  {journal} {Phys. Rev. B}\ }\textbf {\bibinfo {volume}
  {69}},\ \bibinfo {pages} {214423} (\bibinfo {year} {2004})}\BibitemShut
  {NoStop}%
\bibitem [{\citenamefont {Lynn}\ \emph {et~al.}(1976)\citenamefont {Lynn},
  \citenamefont {Shirane},\ and\ \citenamefont {Blume}}]{S_Irmff}%
  \BibitemOpen
  \bibfield  {author} {\bibinfo {author} {\bibfnamefont {J.~W.}\ \bibnamefont
  {Lynn}}, \bibinfo {author} {\bibfnamefont {G.}~\bibnamefont {Shirane}}, \
  and\ \bibinfo {author} {\bibfnamefont {M.}~\bibnamefont {Blume}},\
  }\href@noop {} {\bibfield  {journal} {\bibinfo  {journal} {Phys. Rev. Lett.}\
  }\textbf {\bibinfo {volume} {37}},\ \bibinfo {pages} {154} (\bibinfo {year}
  {1976})}\BibitemShut {NoStop}%
\bibitem [{\citenamefont {Goddard}\ \emph {et~al.}(2008)\citenamefont
  {Goddard}, \citenamefont {Singleton}, \citenamefont {Sengupta}, \citenamefont
  {McDonald}, \citenamefont {Lancaster}, \citenamefont {Blundell},
  \citenamefont {Pratt}, \citenamefont {Cox}, \citenamefont {Harrison},
  \citenamefont {Manson}, \citenamefont {Southerland},\ and\ \citenamefont
  {Schlueter}}]{Goddard2008}%
  \BibitemOpen
  \bibfield  {author} {\bibinfo {author} {\bibfnamefont {P.~A.}\ \bibnamefont
  {Goddard}}, \bibinfo {author} {\bibfnamefont {J.}~\bibnamefont {Singleton}},
  \bibinfo {author} {\bibfnamefont {P.}~\bibnamefont {Sengupta}}, \bibinfo
  {author} {\bibfnamefont {R.~D.}\ \bibnamefont {McDonald}}, \bibinfo {author}
  {\bibfnamefont {T.}~\bibnamefont {Lancaster}}, \bibinfo {author}
  {\bibfnamefont {S.~J.}\ \bibnamefont {Blundell}}, \bibinfo {author}
  {\bibfnamefont {F.~L.}\ \bibnamefont {Pratt}}, \bibinfo {author}
  {\bibfnamefont {S.}~\bibnamefont {Cox}}, \bibinfo {author} {\bibfnamefont
  {N.}~\bibnamefont {Harrison}}, \bibinfo {author} {\bibfnamefont {J.~L.}\
  \bibnamefont {Manson}}, \bibinfo {author} {\bibfnamefont {H.~I.}\
  \bibnamefont {Southerland}}, \ and\ \bibinfo {author} {\bibfnamefont {J.~A.}\
  \bibnamefont {Schlueter}},\ }\href@noop {} {\bibfield  {journal} {\bibinfo
  {journal} {New J. Phys.}\ }\textbf {\bibinfo {volume} {10}},\ \bibinfo
  {pages} {083025} (\bibinfo {year} {2008})}\BibitemShut {NoStop}%
\bibitem [{\citenamefont {Jaime}\ \emph {et~al.}(2006)\citenamefont {Jaime},
  \citenamefont {Lacerda}, \citenamefont {Takano},\ and\ \citenamefont
  {Boebinger}}]{Jaime2006}%
  \BibitemOpen
  \bibfield  {author} {\bibinfo {author} {\bibfnamefont {M.}~\bibnamefont
  {Jaime}}, \bibinfo {author} {\bibfnamefont {A.}~\bibnamefont {Lacerda}},
  \bibinfo {author} {\bibfnamefont {Y.}~\bibnamefont {Takano}}, \ and\ \bibinfo
  {author} {\bibfnamefont {G.~S.}\ \bibnamefont {Boebinger}},\ }\href@noop {}
  {\bibfield  {journal} {\bibinfo  {journal} {J. Phys.: Conf. Ser.}\ }\textbf
  {\bibinfo {volume} {51}},\ \bibinfo {pages} {643} (\bibinfo {year}
  {2006})}\BibitemShut {NoStop}%
\bibitem [{\citenamefont {Nagamiya}\ \emph {et~al.}(1962)\citenamefont
  {Nagamiya}, \citenamefont {Nagata},\ and\ \citenamefont {Kitano}}]{Nagamiya}%
  \BibitemOpen
  \bibfield  {author} {\bibinfo {author} {\bibfnamefont {T.}~\bibnamefont
  {Nagamiya}}, \bibinfo {author} {\bibfnamefont {K.}~\bibnamefont {Nagata}}, \
  and\ \bibinfo {author} {\bibfnamefont {Y.}~\bibnamefont {Kitano}},\
  }\href@noop {} {\bibfield  {journal} {\bibinfo  {journal} {Progr. Theor.
  Phys.}\ }\textbf {\bibinfo {volume} {27}},\ \bibinfo {pages} {1253} (\bibinfo
  {year} {1962})}\BibitemShut {NoStop}%
\bibitem [{\citenamefont {Rousochatzakis}\ and\ \citenamefont
  {Perkins}(2018)}]{Perkins2018}%
  \BibitemOpen
  \bibfield  {author} {\bibinfo {author} {\bibfnamefont {I.}~\bibnamefont
  {Rousochatzakis}}\ and\ \bibinfo {author} {\bibfnamefont {N.~B.}\
  \bibnamefont {Perkins}},\ }\href {\doibase 10.1103/PhysRevB.97.174423}
  {\bibfield  {journal} {\bibinfo  {journal} {Phys. Rev. B}\ }\textbf {\bibinfo
  {volume} {97}},\ \bibinfo {pages} {174423} (\bibinfo {year}
  {2018})}\BibitemShut {NoStop}%
\bibitem [{\citenamefont {Freund}\ \emph {et~al.}(2016)\citenamefont {Freund},
  \citenamefont {Williams}, \citenamefont {Johnson}, \citenamefont {Coldea},
  \citenamefont {Gegenwart},\ and\ \citenamefont {Jesche}}]{Freund}%
  \BibitemOpen
  \bibfield  {author} {\bibinfo {author} {\bibfnamefont {F.}~\bibnamefont
  {Freund}}, \bibinfo {author} {\bibfnamefont {S.~C.}\ \bibnamefont
  {Williams}}, \bibinfo {author} {\bibfnamefont {R.~D.}\ \bibnamefont
  {Johnson}}, \bibinfo {author} {\bibfnamefont {R.}~\bibnamefont {Coldea}},
  \bibinfo {author} {\bibfnamefont {P.}~\bibnamefont {Gegenwart}}, \ and\
  \bibinfo {author} {\bibfnamefont {A.}~\bibnamefont {Jesche}},\ }\href
  {\doibase 10.1038/srep35362} {\bibfield  {journal} {\bibinfo  {journal} {Sci.
  Rep.}\ }\textbf {\bibinfo {volume} {6}},\ \bibinfo {pages} {35362} (\bibinfo
  {year} {2016})}\BibitemShut {NoStop}%
\bibitem [{dat()}]{data_archive}%
  \BibitemOpen
  \href@noop {} {}\bibinfo {note} {Data archive weblink}\BibitemShut {NoStop}%
\end{thebibliography}%

\end{document}